\documentclass[fleqn,usenatbib]{mnras}
\usepackage[T1]{fontenc}
\usepackage[psamsfonts]{amssymb}
\usepackage[dvips]{graphicx}
\usepackage{textcomp}
\usepackage[dvipsnames]{xcolor}
\usepackage{physics}
\usepackage{courier}
\usepackage{natbib}
\usepackage{amsmath,alltt}
\usepackage{nccmath}
\usepackage{tikz}
\usepackage{capt-of}
\usepackage{todonotes}
\usepackage{pdflscape}
\usepackage{multirow}
\usepackage{rotating}
\usepackage{lscape}
\usepackage{subcaption}
\usepackage{array}
\usepackage{mwe}
\usepackage{ae,aecompl}

\title[Chaos and L\'evy Flights in the Three-Body Problem]{Chaos and L\'evy Flights in the Three-Body Problem}

\author[]{Viraj Manwadkar$^{1,3}$, Alessandro A. Trani$^{2}$, Nathan W. C. Leigh$^{3,4}$\\
$^{1}$Department of Astronomy and Astrophysics, University of Chicago, 5640 S. Ellis Ave., Chicago, IL 60637 \\
$^{2}$Department of Astronomy, Graduate School of Science, The University of Tokyo, 7-3-1 Hongo, Bunkyo-ku, Tokyo, 113-0033, Japan \\
$^{3}$Departamento de Astronom\'ia, Facultad de Ciencias F\'isicas y Matem\'aticas, Universidad de Concepci\'on, Chile\\
$^{4}$Department of Astrophysics, American Museum of Natural History, Central Park West and 79th Street, New York, NY 10024}

\date{Accepted 2020 June 17}
\pubyear{2020}
\begin{document}
\label{firstpage}
\pagerange{\pageref{firstpage}--\pageref{lastpage}} 
\maketitle

\begin{abstract}

We study chaos and L\'evy flights in the general gravitational three-body problem. We introduce new metrics to characterize the time evolution and final lifetime distributions, namely Scramble Density $\mathcal{S}$ and the LF index $\mathcal{L}$, that are derived from the Agekyan-Anosova maps and homology radius $R_{\mathcal{H}}$. Based on these metrics, we develop detailed procedures to isolate the ergodic interactions and L\'evy flight interactions. This enables us to study the three-body lifetime distribution in more detail by decomposing it into the individual distributions from the different kinds of interactions. We observe that ergodic interactions follow an exponential decay distribution similar to that of radioactive decay. Meanwhile, L\'evy flight interactions follow a power-law distribution. L\'evy flights in fact dominate the tail of the general three-body lifetime distribution, providing conclusive evidence for the speculated connection between power-law tails and L\'evy flight interactions. We propose a new physically-motivated model for the lifetime distribution of three-body systems and discuss how it can be used to extract information about the underlying ergodic and L\'evy flight interactions.  We discuss ejection probabilities in three-body systems in the ergodic limit and compare it to previous ergodic formalisms. We introduce a novel mechanism for a three-body relaxation process and discuss its relevance in general three-body systems.

\end{abstract}


\section{Introduction} \label{intro}

The three-body problem is one of the longest standing problems in physics, dating back to Newton. Newton provided a closed, analytical solution for the two-body problem. However, the three-body problem has withstood analytical solutions, despite great strides in related areas of physics, mathematics and especially computing (see \citet{valtonen06} for more details). The reason being that the three-body problem is a complex one to understand due to its chaotic quasi-stable nature, as first pointed out by \citet{poincare92}. However, with current computational capabilities, we have been able to provide statistical descriptions of the end states of the general three-body problem (for example see \citet{monaghan76a}, \citet{monaghan76b},\citet{leigh16}, and \citet{stone19}). 

There are mainly 2 different states in which the three-body system can exist during its evolution: a hierarchical state where there is a close binary pair and a temporary single; or a chaotic state where the 3 masses are in approximate energy equipartition, rapidly and chaotically exchanging energy and angular momentum \citep[e.g.][]{anosova85,valtonen91}. The lifetime of the system, also known as the disruption time denoted by $\tau_{D}$, is defined as the time until one of the masses is ejected, that is, has net positive energy. \citet{leigh16} has shown that we can fit the cumulative lifetime distributions of three-body and higher-N systems with a half-life formalism. This is analogous to radioactive decay \citep{ibragimov18}, where the half-life is defined as the time when the probability that 50\% of a sample of interacting three-body systems are still gravitationally bound. In the point-particle limit, every individual three-body interaction ends with the ejection of one particle to spatial infinity because, in the point-particle limit, we do not consider collisions.

This curious connection between these macro self-gravitating systems and micro nuclear systems might prove to be insightful in understanding the underlying physics giving rise to different macroscopic outcome states. For example, one cannot directly observe the time evolution of nuclear systems at the particle-level.  However, we can do this for self-gravitating N-body systems as they evolve in time \citep[e.g.][]{leigh18}. Our understanding of the binding energy of the nucleons in the nucleus and the phenomena of neutron capture is analogous to the capture of bodies in the N-body problem. Fundamentally, the time evolution of the gravitational three-body system is a thermodynamic problem; one studies the diffusion of energy and angular momentum between interacting particles in the system, and how this imprints itself onto the macroscopic parameters of the end-states (see the discussion in \citet{leigh18} for more details), such as the final distribution of interaction lifetimes, the velocities of the ejected particles, the properties of the left-over binary and even which objects will comprise the ejected single and the left-over binary. 

Recently, \citet{shevchenko10}, \citet{orlov10} and \citet{leigh16} showed that the tails of the lifetime distributions of three-body systems are algebraic and not exponential like the initial part of the distribution. The algebraic nature of the distribution tails is because of the existence of L\'evy flights as discussed in \citet{shevchenko10} and \citet{orlov10}. However, the precise physical origins and the nature of these L\'evy flights is still not well understood (see the discussions in \citet{leigh16} and \citet{ibragimov18} for more details). 

The key idea behind this paper is to develop and test a new quantitative parameter to describe the degree of chaos developed in a self-gravitating system of point particles. Furthermore, we continue the study of the algebraic tails of the lifetime distributions and take a closer look at the physical origins of these L\'evy flights \citep{leigh16,ibragimov18}. We also aim to study the half-lives and ejection probabilities of three-body systems in the ergodic limit.

In Section~\ref{metrics} we present and discuss the metrics, namely Scramble Density and LF Index, to quantify chaos and the L\'evy flight nature respectively. In Section~\ref{sec:methods} we present our numerical scattering experiments, and describe our fitting procedure for obtaining characteristic half-lives and power-law indices for the cumulative lifetime distributions. We present our results in Section~\ref{results}, and discuss their significance for chaos, L\'evy flights and general lifetime distributions in three-body systems in Section~\ref{discussion}. We summarize our main conclusions in Section~\ref{summary}.

\section{Quantifying Chaos} \label{metrics}

In Section~\ref{construct}, we introduce the Agekyan-Anosova mapping system which is useful for analyzing three-body interactions and their various types. In Section~\ref{sssec:levy}, we discuss the L\'evy flight nature of the three-body interactions and introduce a new metric to quantify this L\'evy flight behavior. In Section~\ref{sssec:scram}, we introduce a new metric for chaos called 'Scramble Density'. We apply each of these metrics to study the lifetime distribution statistics and general three-body system properties in Section~\ref{results}.  

Throughout the paper, we denote the disruption time/lifetime of a system by $\tau_{D}$ where $\tau_{D}$ is in units of crossing-times $\tau\textsubscript{cr}$ where $\tau\textsubscript{cr} = \frac{GM^{5/2}}{(2E_0)^{3/2}}$ where $M$ is the total mass of the system and $E_0$ is the total energy of the system. (\citet{valtonen06}, \citet{leigh16}). For reference, Table~\ref{table:tcr} contains the $\tau\textsubscript{cr}$ in years for the 3 systems under consideration. As a result, the lifetimes of the system that are reported throughout this paper are unit-less quantities. 

\begin{table}
\caption{$\tau\textsubscript{cr}$ in years for the 3 three-body systems under consideration.}
\label{table:tcr}
\centering
\begin{tabular}{cc}
\hline
 Masses($M\textsubscript{\(\odot\)}$) & $\tau\textsubscript{cr}$ (yrs) \\
\hline
15,15,15 & $48.379$ \\
12.5,15,17.5 & $40.675$ \\
10,15,20 & $35.010$ \\
\hline
\end{tabular}
\end{table}

\subsection{Agekyan-Anosova Maps}\label{construct}

\citet{agekyan67} developed a mapping system known as the Agekyan-Anosova map (or AA map) for analyzing the geometrical properties of a co-planar three-body system over the course of its time evolution. The location of a dot on the AA map represents the shape of the triangle formed by the three masses in space at a given instant of time. The AA map is constructed on the Cartesian plane by fixing the most distant two particles at $(0.5, 0)$ and $(-0.5,0)$, so that a dot indicates the location of the third body in this re-scaled reference frame. The location of the dot does not depend on any other parameters like mass or velocity. All possible spatial configurations of the three-body problem are contained in the region between $(0,0)$, $(0.5,0)$ and $(0,\sqrt{3}/2)$, which is also called as the $\mathcal{D}$ region. The AA map is well-defined for co-planar three-body interactions while for general 3D three-body interactions, the AA map is not well-defined.

\begin{figure}
\includegraphics[width=\columnwidth]{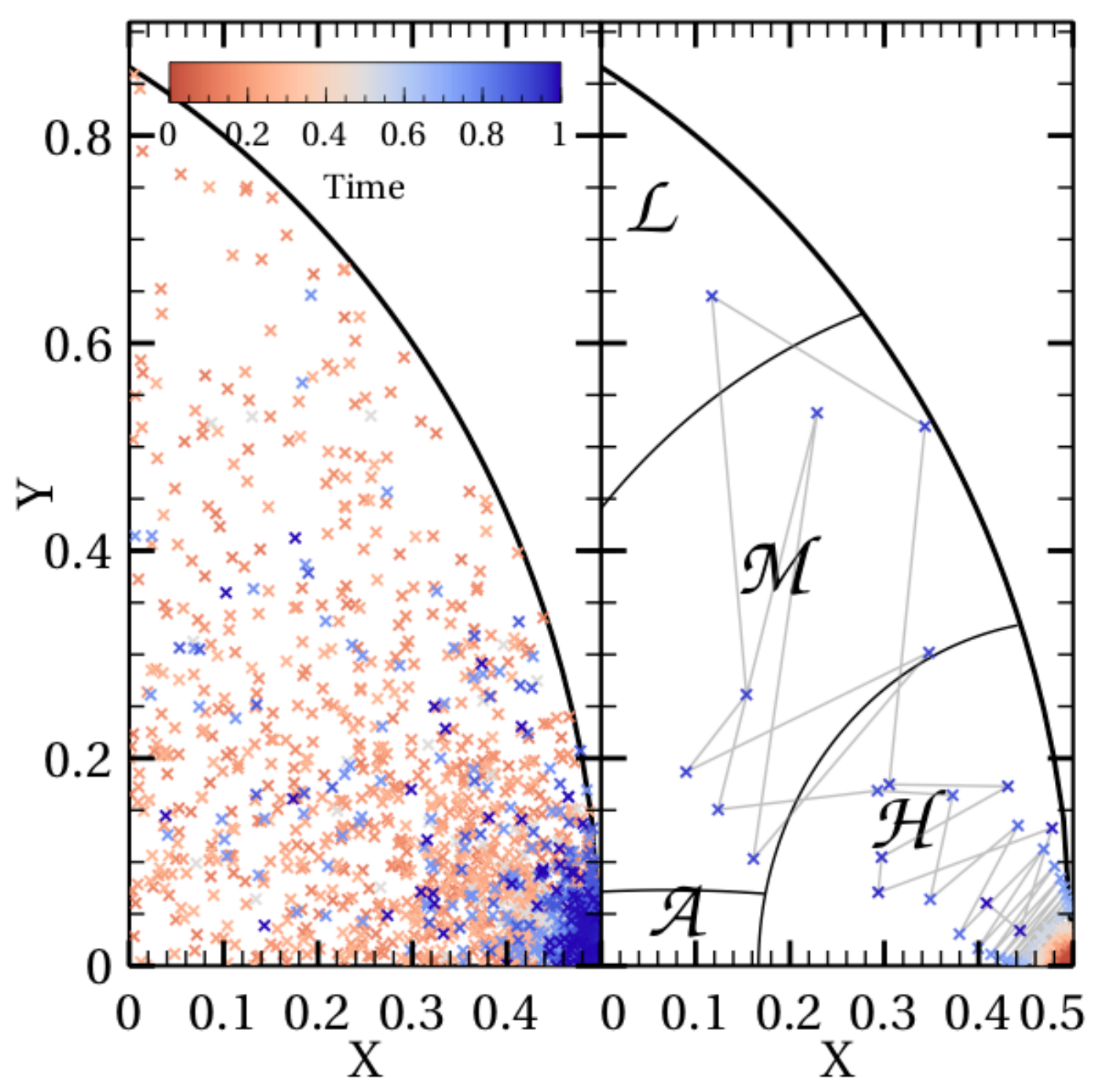}
\caption[Agekyan-Anosova Map]{The Agekyan-Anosova homology map. Left and right-hand panels display the evolution of two distinct coplanar, equal-mass simulations with the setup described in the Section~\ref{sec:sims}. The color-code represents the evolution in time. Left-hand panel: long-lived triple ($\tau_{D}=4.98$). Right-hand panel: prompt-interaction ($\tau_{D}=0.537$). For details about the construction of the AA map, see \citet{heinamaki98}.}
\label{fig:fig1}
\end{figure}

As \citet{heinamaki98} discussed, the $\mathcal{D}$ region in the AA map can be split into 4 characteristic areas. The region in the top corner is called the Lagrangian ($\mathcal{L}$) area, referring to the well-known family of solutions of the three-body problem forming equilateral triangles. The region in the bottom right corner is called the Hierarchical ($\mathcal{H}$) area and represents systems with a hierarchical structure, that is, a distinct binary pair and single mass configuration exists. The region in the bottom is called the Alignment ($\mathcal{A}$) area and represents triangles for which the 3 masses are roughly in a linear configuration. The remaining area is simply called the Middle ($\mathcal{M}$) area. 

The relative positions of the three masses can be inferred based on the location of the dot in the AA map. For example, if we have a hierarchical triple composed of a compact inner binary pair and a single on a wide orbit, we will have a dot in the $\mathcal{H}$ region. Therefore, the most stable three-body systems lie in the $\mathcal{H}$ region. Furthermore, during chaotic interactions, the position of the dot in the AA map will change rapidly in an unpredictable manner. 

In Figure~\ref{fig:fig1} we show the evolution of two simulations of three-body interactions in the AA map. A long-lived chaotic interaction is show in the left-hand panel, while a short lived one is shown on the right-hand panel. Both simulations begin and end in the $\mathcal H$ region, which indicates that the system is forming a hierarchical triple at the end of the interaction. In the long-lived interaction, the triple undergoes several democratic interactions (located in the $\mathcal L$ region) and excursions. Excursions are defined as temporary hierarchical configurations (located in the $\mathcal H$ region) of the three-body system. The short-lived simulation passes by the $\mathcal L$ region only once before breaking up, indicating that the system undergoes a single strong interaction.

\begin{figure*}
\centering
\begin{subfigure}[b]{0.475\textwidth}
\centering
\includegraphics[width=\textwidth]{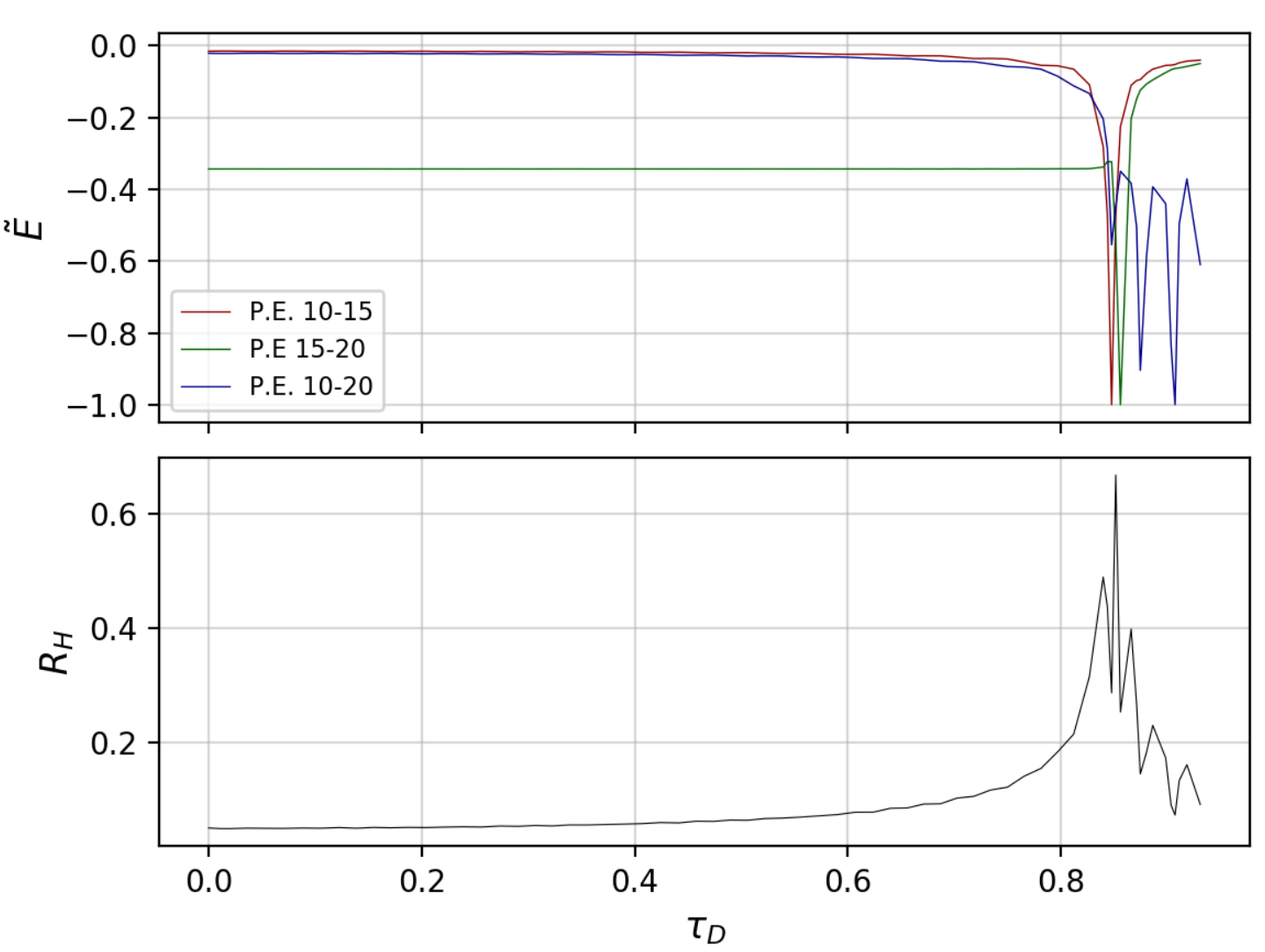}
\caption[]%
{{\small Prompt Three-Body Interaction}}
\label{fig:prompt}
\end{subfigure}
\hfill
\begin{subfigure}[b]{0.475\textwidth}
\centering 
\includegraphics[width=\textwidth]{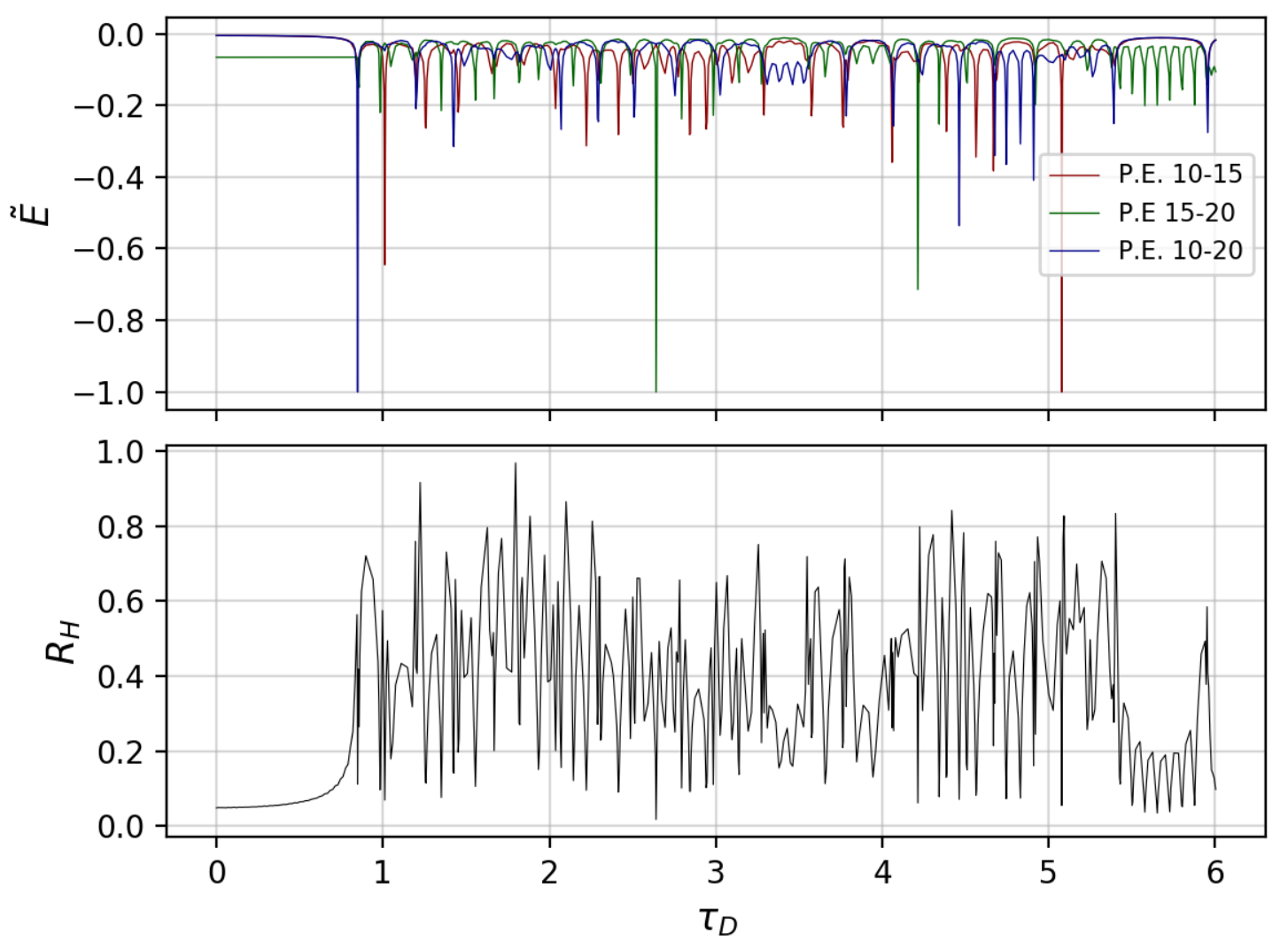}
\caption[]%
{{\small Chaotic Three-Body Interaction}}    
\label{fig:chaos_homo}
\end{subfigure}
\vskip\baselineskip
\begin{subfigure}[b]{0.475\textwidth}   
\centering 
\includegraphics[width=\textwidth]{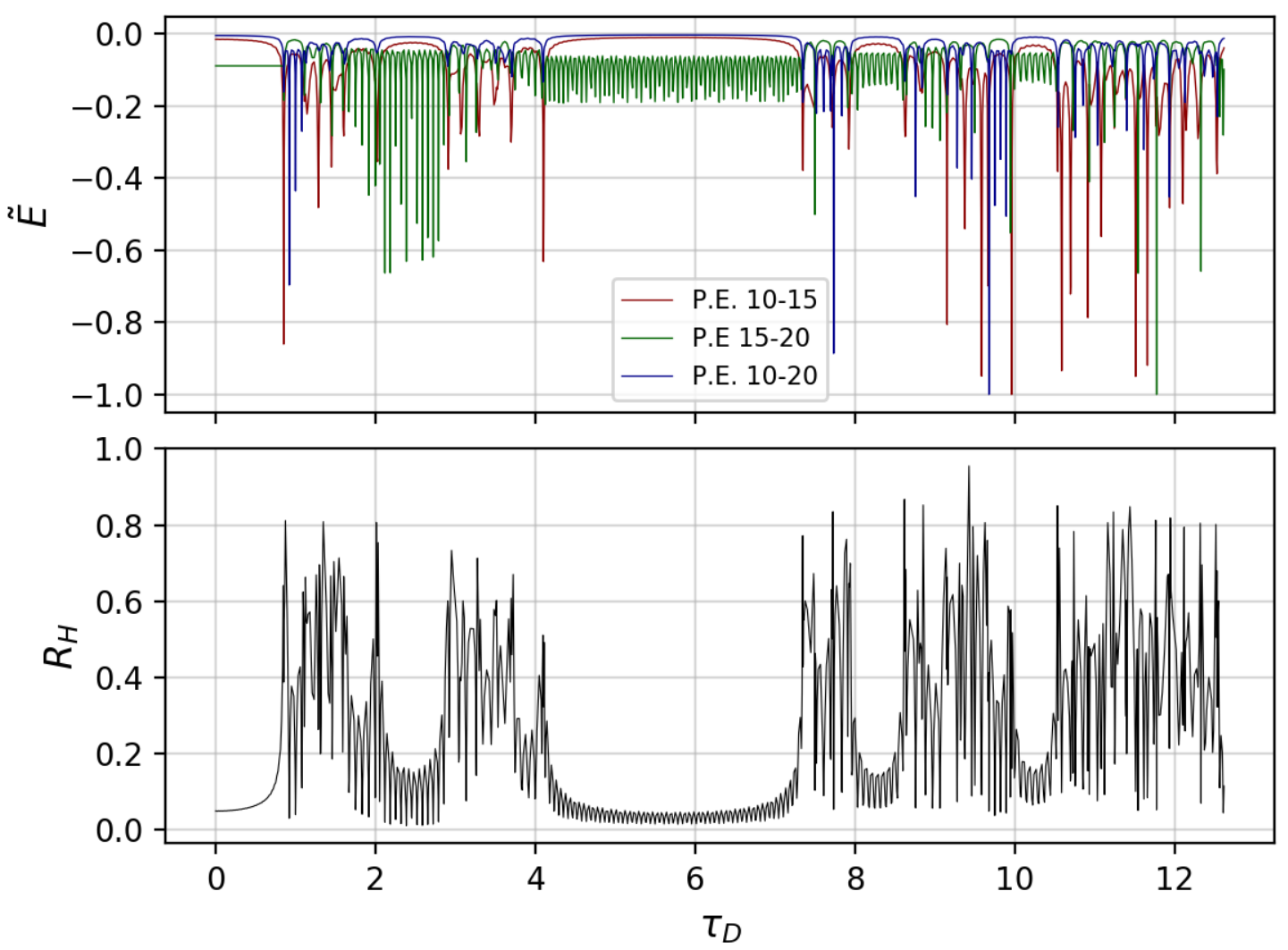}
\caption[]%
{{\small Intermittent Three-Body behavior }}
\label{fig:mix}
\end{subfigure}
\hfill
\begin{subfigure}[b]{0.475\textwidth}  
\centering 
\includegraphics[width=\textwidth]{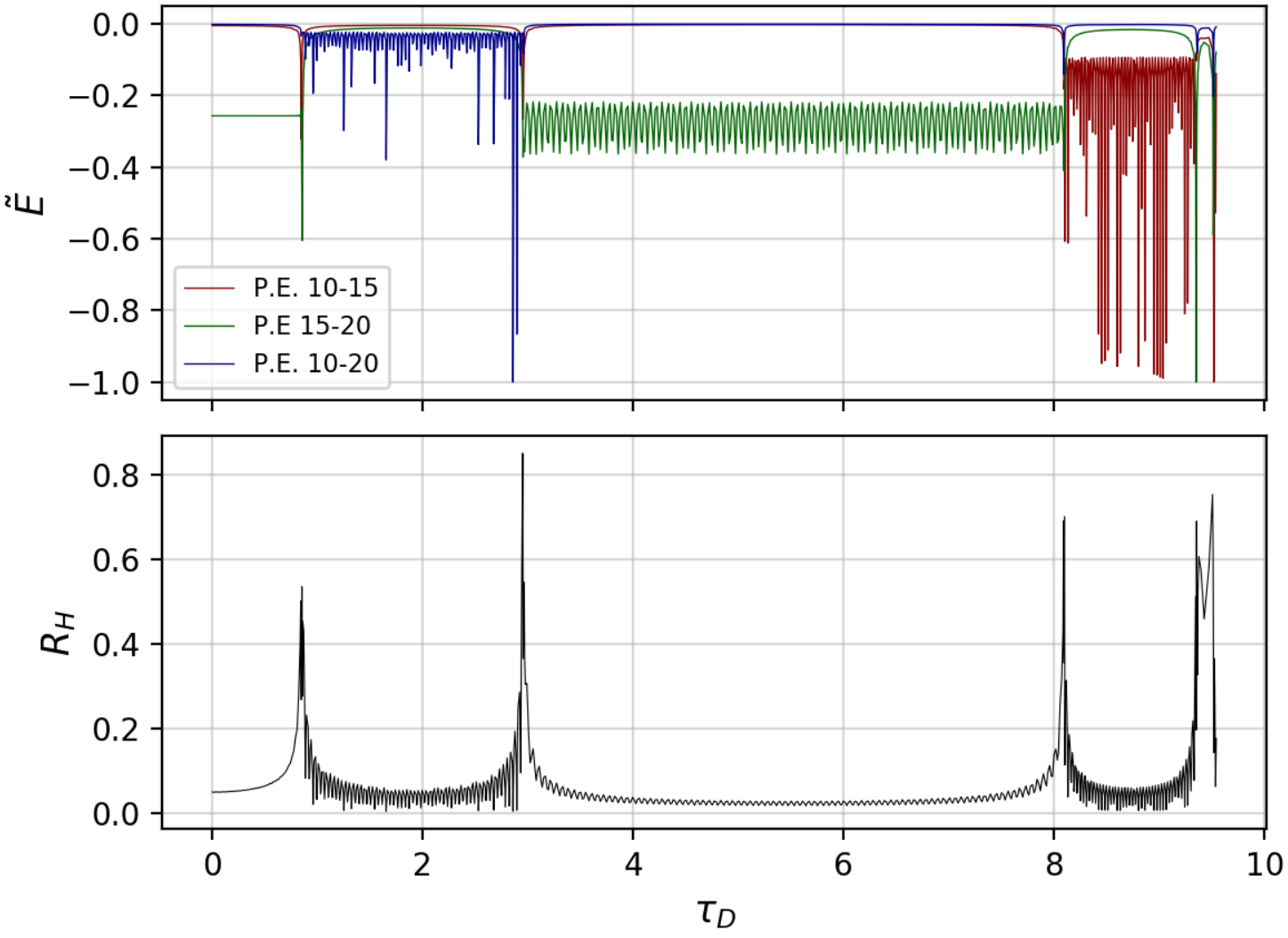}
\caption[]%
{{\small LF2-type interaction}}    
\label{fig:levy}
\end{subfigure}
\caption{Homology Radius profiles for different kinds of three-body interactions}
\end{figure*}

\subsection{Intermittency and the Homology Radius} \label{exp}

Intermittency is the phenomenon where stable or periodic motion in a system is abruptly disrupted by random bursts of chaos. One can study intermittency in three-body dynamics by studying the homology mapping of three-body systems in the AA map. The term ``homology mapping'' and ``homology radius'' was first introduced in \citet{heinamaki98}. This mapping system was developed to study the ``homological'' properties, that is, the geometrical properties of the triangle formed by the three bodies in the system, hence the name `homology mapping'. We present a revised version of the homology radius here. We define the homology radius, denoted by $R_{\mathcal{H}}$, as the distance between $(0.5, 0)$ and the dot in the $\mathcal{D}$ region of the AA map in Figure~\ref{fig:fig1}. Now, as said in Section~\ref{construct}, the AA map is only well-defined for co-planar three-body interactions. However, the homology radius can work for any configuration in 3D space. Therefore, a more general definition of homology radius $R_{\mathcal{H}}$ is the normalized shortest distance between any 2 bodies in the three-body system. \footnote{By normalized distance, we mean that distances scaled in a manner such that the longest distance between any 2 bodies in three-body system at a given instant of time is unit length.} Note that the homology radius is not a quantity associated with each body in the three-body system; the homology radius describes the entire system. Defining the homology radius this way is ideal for studying chaos; when $R_{\mathcal{H}} > 0.33$, the dot has moved outside the stable $\mathcal{H}$-region. 

Figure 2 contains the homology radius ($R_\mathcal{H}$) as a function of time for different kinds of three-body interactions in the 10,15,20$M\textsubscript{\(\odot\)}$ system. We will call such plots as homology radius profiles. Along with the homology radius profiles, the normalized potential energies between the 3 masses are also shown to emphasize the physical similarities between $R_\mathcal{H}$ and physical state of the system. Looking at the homology radius profiles in Figure 2, one observes mainly 2 features. When the system is in a hierarchical state, that is when $R_{\mathcal{H}} < 0.33$, the radius is undergoing constant period, small amplitude oscillations. In fact the period of the homology radius oscillations in this stable state corresponds to the period of the binary pair. On the other hand, when the system is undergoing a chaotic interplay, that is the system is no longer hierarchical, we see violent, rapid, large amplitude oscillations in the homology radius. In terms of the AA map, this corresponds to the dot moving rapidly in a chaotic manner throughout the $\mathcal{D}$ region. Therefore, when the single comes back from its excursion, the dot moves outside the $\mathcal{H}$-region; that is, the homology radius is more than 0.33, and a chaotic interplay often starts. For all the interactions in our case, we are initially starting with a hierarchical structure, therefore, our homology radius starts out small. 

As elucidated above, the homology radius profile of a three-body interaction gives a lot of information about the nature of the interaction. It can be used to understand, quantify and characterize the specifics of the internal evolution of the system as it unfolds. For example, in Figure~\ref{fig:prompt}, after the initial drop of the single, we see a short spike in $R_{\mathcal{H}}$ and the immediate ejection of one of the masses. These spikes in the homology radius correspond to when the single drops into the binary pair resulting in a state of no hierarchical structure. This is when the chaotic interplay occurs, and all three particles are in a state of approximate energy equipartition. We define these spikes as `scrambles' and will discuss them in more detail in Section~\ref{sssec:scram}.

In Figure~\ref{fig:mix} and Figure~\ref{fig:chaos_homo}, we see a predominantly chaotic interaction because the evolution of the system is dominated by these chaotic spikes. However, in Figure~\ref{fig:mix}, we see islands of stability interspersed within the chaos in the evolution. One can identify these islands of stability by the constant period and constant amplitude oscillations in the homology radius. In Figure~\ref{fig:levy}, we observe long excursions that dominate the entire evolution of the system. These long flights are called as L\'evy flight which we will discuss more in the following section.

\subsection{L\'evy Flights}
\label{sssec:levy}

L\'evy flights are a type of random walk motion that have a heavy tailed distribution. In case of L\'evy flights in the three-body system, the random walks are just one-sided increments in the orbital period of the body. Therefore, they lead to a heavy-tailed lifetime distribution of three-body systems.

\citet{shevchenko10} studied L\'evy flights in the hierarchical restricted three-body problem. In this specialized case of the three-body problem, the mass of the tertiary is insignificant compared to the other 2 masses in the system. It is also assumed in his study that the orbital size of the tertiary is much larger than the orbital size of the binary. That is, only weak perturbations of the binary center of mass by the tertiary are considered as the impact parameter is always large. Such systems are described by the Kepler map which was developed by \citet{petrosky86} and \citet{chirikov89} and is used, for example, to study the chaotic behavior of comets \citep{shevchenko11}. However, it is important to remember that we are considering the general three-body system where we do allow for strong interactions and also there is a more democratic distribution of masses in our three-body systems. This will lead to different physical phenomena and kinds of interactions. None the less, the study of \citet{shevchenko10} will give us insight into the L\'evy flight nature of the general three-body problem, so we consider it here. 

In the hierarchical restricted three-body system, \citet{shevchenko10} identifies 2 different kinds of Hamiltonian intermittency and corresponding kinds of L\'evy flights. The first kind of Hamiltonian intermittency occurs when the trajectory is ``stochastized'' by encounters with the separatix in the Kepler map, where the separatix is a boundary separating bound motion from unbound motion. The second kind of intermittency takes place when a trajectory sticks to the ``chaos-order'' boundary present in phase space. The three-body phase space is what is called a ``divided phase space''; that is, the phase space is divided into regions of ordered and chaotic motion separated by highly intricate borders \citep{boris83}. During these sticking events, one observes series of excursions that are interrupted by democratic states in which all three particles have comparable energies and are confined to a small volume. L\'evy flights occurring due to the first kind of Hamiltonian intermittency are called L\'evy flights of the first kind (LF1-type). LF1-type interactions appear as sudden jumps in orbital size and period; that is, a sudden long excursion that dominates the time evolution of the interaction. On the other hand, L\'evy flights occurring due to the second kind of Hamiltonian intermittency are called L\'evy flights of the second kind (LF2-type). LF2-type interactions appear as a series of excursions of similar orbital size and eccentricity. For more information about these 2 kinds of L\'evy flights, refer to \citet{shevchenko10}. We consider these 2 kinds of L\'evy flights with a word of caution in our problem even though there are connections that are insightful. 

A key property of L\'evy flights in general is that they are characterized by substantial time intervals (excursions) where the system has hierarchical structure. Similar behavior is also observed in the homology radius profiles of our system (see, for instance, Figure~\ref{fig:levy}). We will discuss in more detail in Section~\ref{levy_types} the relevance of these 2 kinds of L\'evy flights to the general three-body problem. 

The goal now is to establish some quantitative measures to identify L\'evy flight interactions. Inspired from the features of these kinds of motion that one can see in the homology radius profile in Figure~\ref{fig:levy}, we introduce a new parameter called the LF index. We define it as

\begin{ceqn}
\begin{align}
\mathcal{L} = \frac{\tau_{H}}{\tau_{D}}
\end{align}
\end{ceqn}

where $\tau_{D}$ is the lifetime of the system and $\tau_{H}$ is the total duration the system spends in the hierarchical regime. As mentioned earlier, all units of time are in units of crossing time $\tau\textsubscript{cr}$. $\tau_{H}$ is essentially the sum of the duration of all the excursions during the evolution. The value of $\mathcal{L}$ ranges between 0 and 1 where it is close to 1 for L\'evy flight interactions. Note that the $\mathcal{L}$ is a unit-less parameter. This index, though simple in construction, will be crucial in understanding the power-law tail of the cumulative lifetime distribution and the L\'evy flight nature of the three-body problem.

\subsection{Scramble Density}
\label{sssec:scram}

As discussed in \citet{leigh16b}, \citet{leigh18} and \citet{stone19}, a ``scramble'' in a three-body system is when no pairwise binaries exist; that is, no hierarchical structure is present. Formally, a scramble occurs every time the homology radius peaks above $R_{\mathcal{H}} = 0.33$. According to \citet{stone19}, scrambles are the key dynamical state that ergodicizes a three-body system. \citet{stone19} used the number of scrambles $N_{\rm T}$ to define the point at which individual simulations of the chaotic three-body problem enter the ergodic regime (i.e., the system has "forgotten" all memory of its initial conditions, except for the particle masses, total energy and total momentum).  Specifically,  for a given total encounter energy and momentum, the authors found that the three-body systems they considered had predominantly entered the ergodic regime when $N_{\rm T} >$ 2. Therefore, based on this, we introduce a new quantitative index called the `Scramble Density` to measure chaos in three-body systems. 

\begin{table*}
\caption{Table of all the three-body simulation sets we perform with their respective initial parameters.}
\label{table:tab1}
\centering
\begin{tabular}{p{1cm}p{1cm}p{1cm}p{1cm}p{1cm}p{1.5cm}p{1cm}p{1.5cm}}
\hline
Label & No. of Simulations &  Binary Pair ($M\textsubscript{\(\odot\)}$) & Single ($M\textsubscript{\(\odot\)}$) & Semi-major axis (AU) & Eccentricity & Binary Phase & Inclination \\
\hline
A&$10^6$&15,15&15&5&0&[0,2$\pi$]&[0,$\pi$]\\
B&$10^6$&15,17.5&12.5&5&0&[0,2$\pi$]&[0,$\pi$]\\
C&$10^6$&15,20&10&5&0&[0,2$\pi$]&[0,$\pi$]\\
\hline
\end{tabular}
\end{table*}

The Scramble Density (denoted by $\mathcal{S}$) is defined as

\begin{ceqn}
\begin{align}
   \mathcal{S} = \frac{N_{T}}{\tau_{D}}
\end{align}
\end{ceqn}

where $N_{T}$ is the total number of scrambles in the lifetime of the system and $\tau_{D}$ is the disruption time of the system in crossing units $\tau_{cr}$. This parameter is designed such that the value of $\mathcal{S}$ is high for ergodic interactions and low for non-ergodic interactions. The exact quantification of what is high and low will be shown in Section~\ref{sssec:scram_results}. Note that the Scramble Density is a unit-less parameter.

The reason why $\mathcal{S}$ is high for ergodic interactions is that the larger the number of scrambles over the course of the lifetime of a given three-body interaction, the more opportunity it has to entire the ergodic regime. One could ask: why not just define the index as being equal to the number of scrambles? The reason is that this index helps differentiate between ergodic interactions and L\'evy flight interactions. It is possible for both of these interactions to have, for example, 50 scrambles, but they have a drastically different evolution and $\tau_{D}$. $\mathcal{S}$ helps to differentiate between these two types of interactions by giving a low value of $\mathcal{S}$ for a L\'evy flight interaction, and a higher value for the ergodic interactions. 

We will discuss in Section~\ref{results} regarding how to isolate these ergodic interactions, which we call EC-type interactions. ``EC-type'' stands for `exponential cumulative type'. This name derives from a distinctive feature of ergodic interactions that their cumulative lifetime distribution follows an exponential distribution. We will show that this subset of the total lifetime distribution corresponds to the initial exponential drop in the cumulative lifetime distributions. This will be discussed in detail in Section~\ref{results}. 


\section{Methods} \label{sec:methods}

In Section~\ref{sec:sims}, we present the numerical scattering experiments used in this paper to address the over-arching questions identified in Section~\ref{intro}. In Section~\ref{sec:model_fitting}, we present and discuss our methodology in Section for fitting the cumulative lifetime distributions, and for extracting corresponding half-lives or power-law indices.

\subsection{Numerical scattering experiments}\label{sec:sims}

We use the TSUNAMI N-body code to integrate the time evolution of a series of single-binary encounters. TSUNAMI is an implementation of Mikkola's algorithmic regularization. The TSUNAMI code uses a combination of numerical techniques to improve the accuracy of the integration, which makes it particularly suitable to simulate strong dynamical interactions with high mass ratios and hierarchical architectures. The core of the TSUNAMI code is constituted by a Bulirsch-Stoer extrapolation scheme (\citet{stoer80}) on top of a second-order Leapfrog algorithm, derived from a time transformed Hamiltonian (\citealt{mikkola99a}, \citealt{mikkola99b}). We also implement optional higher order symplectic schemes \citep{yoshida90} without the Bulirsch-Stoer. Additionally, TSUNAMI includes velocity-dependent forces, namely the post-Newtonian terms 1PN, 2PN and 2.5PN from \citet{blanchet14}, the tidal drag-force from \citet{samsing18}  and the equilibrium tide from \citet{hut81}. However, in the present work, we do not consider any additional forces besides the purely Newtonian ones. We also neglect collisions between the particles. More details on the code will be presented in a following work (Trani et. al, in preparation).

To determine the state of the three-body system, we implement a simple classification scheme based on energy and stability criteria. The hierarchy state of the triple is checked at every timestep: we consider the most bound pair as a binary, and check if the third body is bound to the binary centre-of-mass. If the third body is unbound, we check if the binary-single pair is converging or diverging on a hyperbolic orbit: if diverging, we record the breakup time $t_\mathrm{break}$ and stop the simulation when the single is 10 times the binary semi-major axis away from the binary centre of mass; if converging, we estimate the time of closest approach and continue the simulation. If the binary-single forms a bound pair, we check its stability using the criterion of \citet{mardling01}: if unstable, we estimate the next pericenter approach and continue the simulation; if instead the system forms a stable hierarchical triple, we wait for 10 orbits of the outer binary and then stop the simulation. If no bound pairs exist, we check if each particle is bound to the centre of mass. If no particle is bound to the centre of mass, we classify the end state as a triple breakup. Note that both a triple breakup and the formation of a stable triple are forbidden with the initial conditions considered in this work \citep{hut83}.

\begin{figure*}
\includegraphics[width=\textwidth]{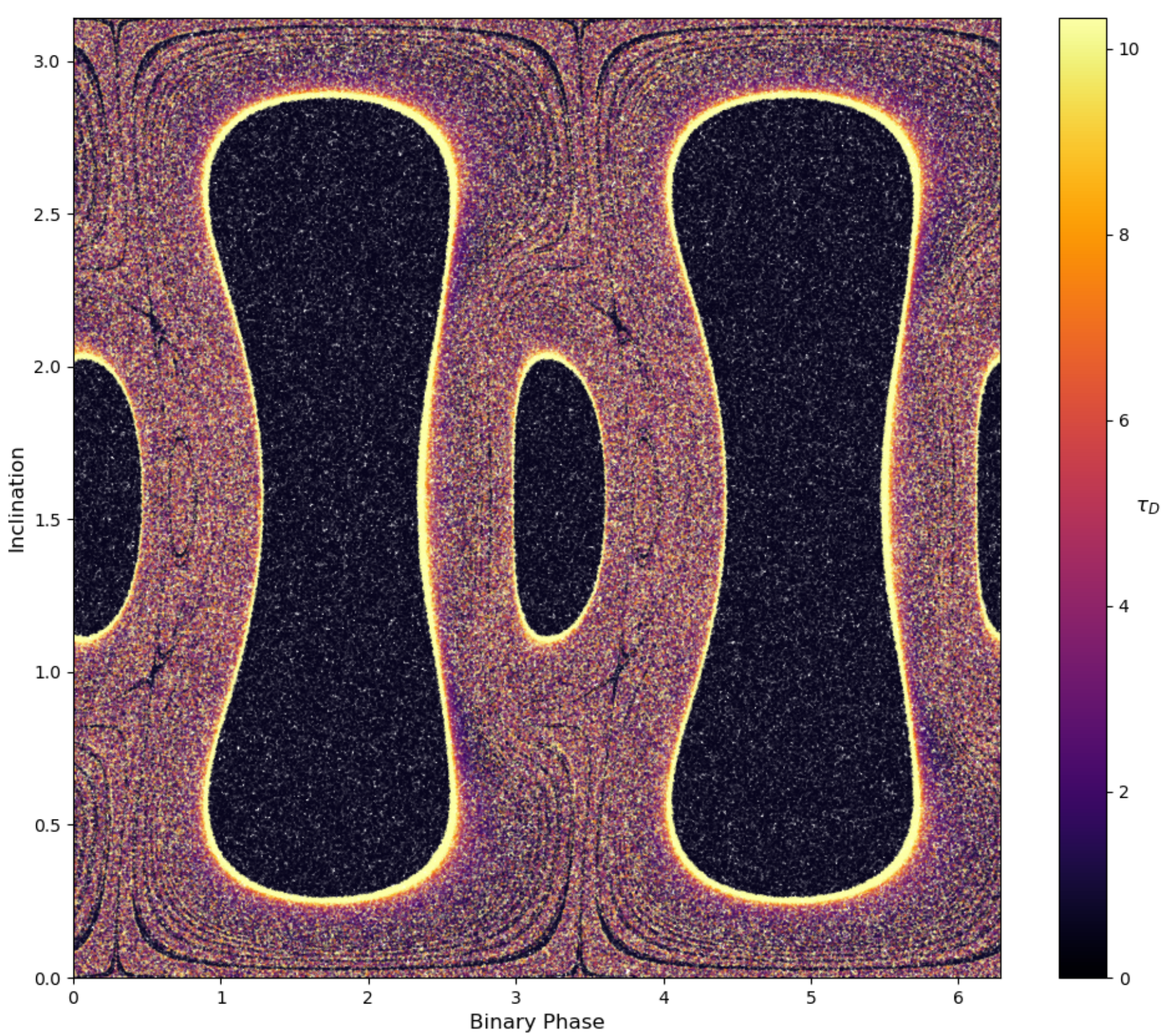}
\caption{Phase map color coded according to lifetime $\tau_{D}$ for the three-body system with masses 15,15 and 15 $M\textsubscript{\(\odot\)}$. The white dots in the plot are just because of the discrete nature of the inclination and binary phase values of our ensemble and have no physical meaning.}
\label{fig:fig3}
\end{figure*}

In this present work, we consider 3 sets with different particle masses; each set comprises $10^6$ realizations. The mass distribution we choose are the equal mass case $15,15,15 M\textsubscript{\(\odot\)}$ (set~A), $12.5,15,17 M\textsubscript{\(\odot\)}$ (set~B) and $10,15,20 M\textsubscript{\(\odot\)}$ (set~C). In all sets, the initial binary has $a_{\rm i} =$ 5 AU, and eccentricity $e_{\rm i} =$ 0. 
The initial distance of the single from the binary is 100 AU. The impact parameter for all simulations is fixed at $b = 0$. The centre of mass of the binary and the single are initially at rest. In each of these sets, the binary is composed of the heavier 2 masses with the single being the smallest mass. In each realization, the true anomaly of the binary is drawn uniformly between 0 and 2$\pi$, while the inclination against the line joining the binary centre of mass and the single is uniform between 0 and $\pi$.

\begin{figure*}
    \centering
        \begin{subfigure}[b]{0.485\textwidth}
            \centering
            \includegraphics[width=\textwidth]{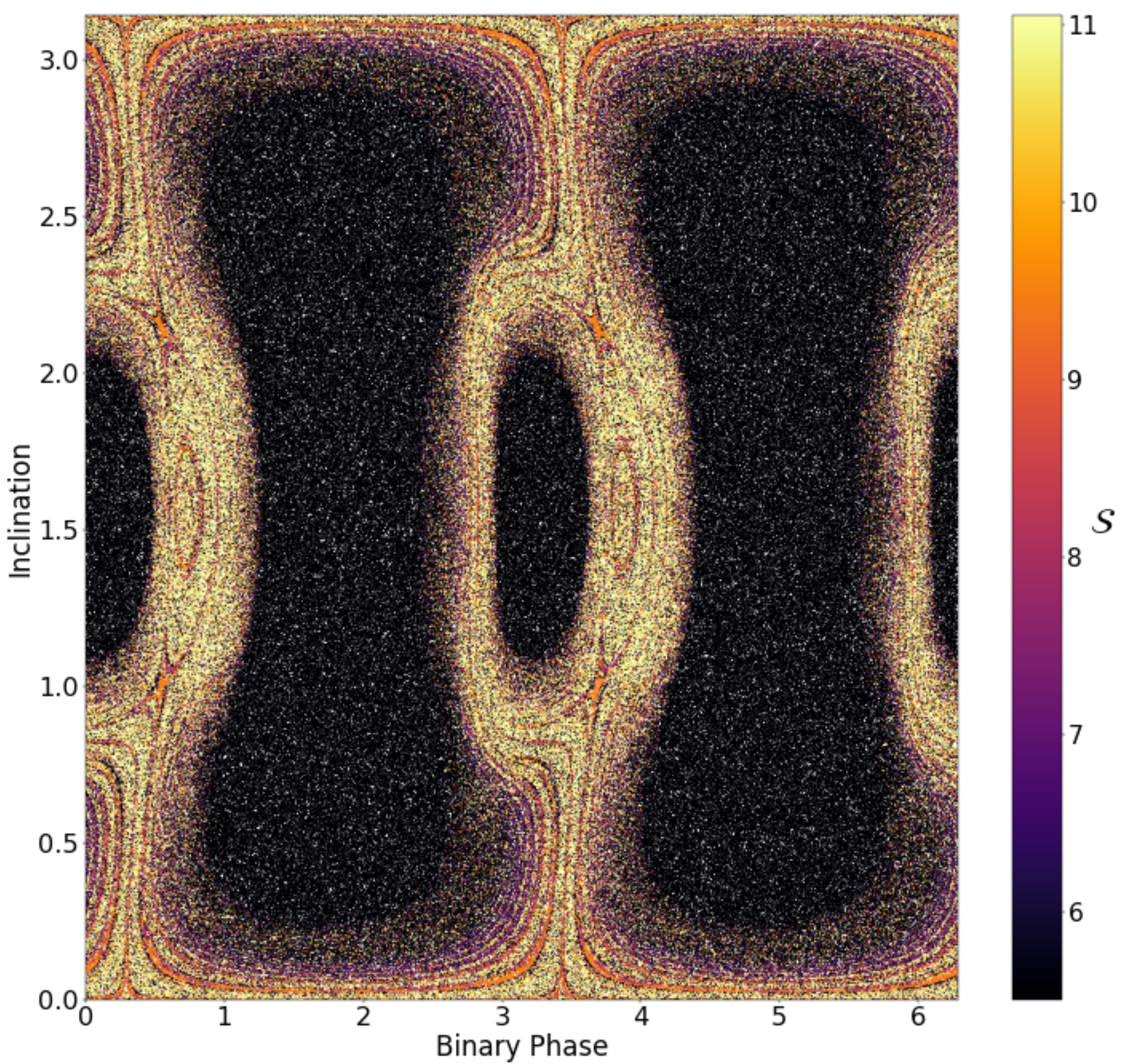}
            \caption[]%
            {{\small Phase map color coded according to Scramble Density $\mathcal{S}$ for the 15,15,15 $M\textsubscript{\(\odot\)}$ system.}}    
            \label{fig:scram_index_phase}
        \end{subfigure}
        \hfill
        \begin{subfigure}[b]{0.485\textwidth}  
            \centering 
            \includegraphics[width=\textwidth]{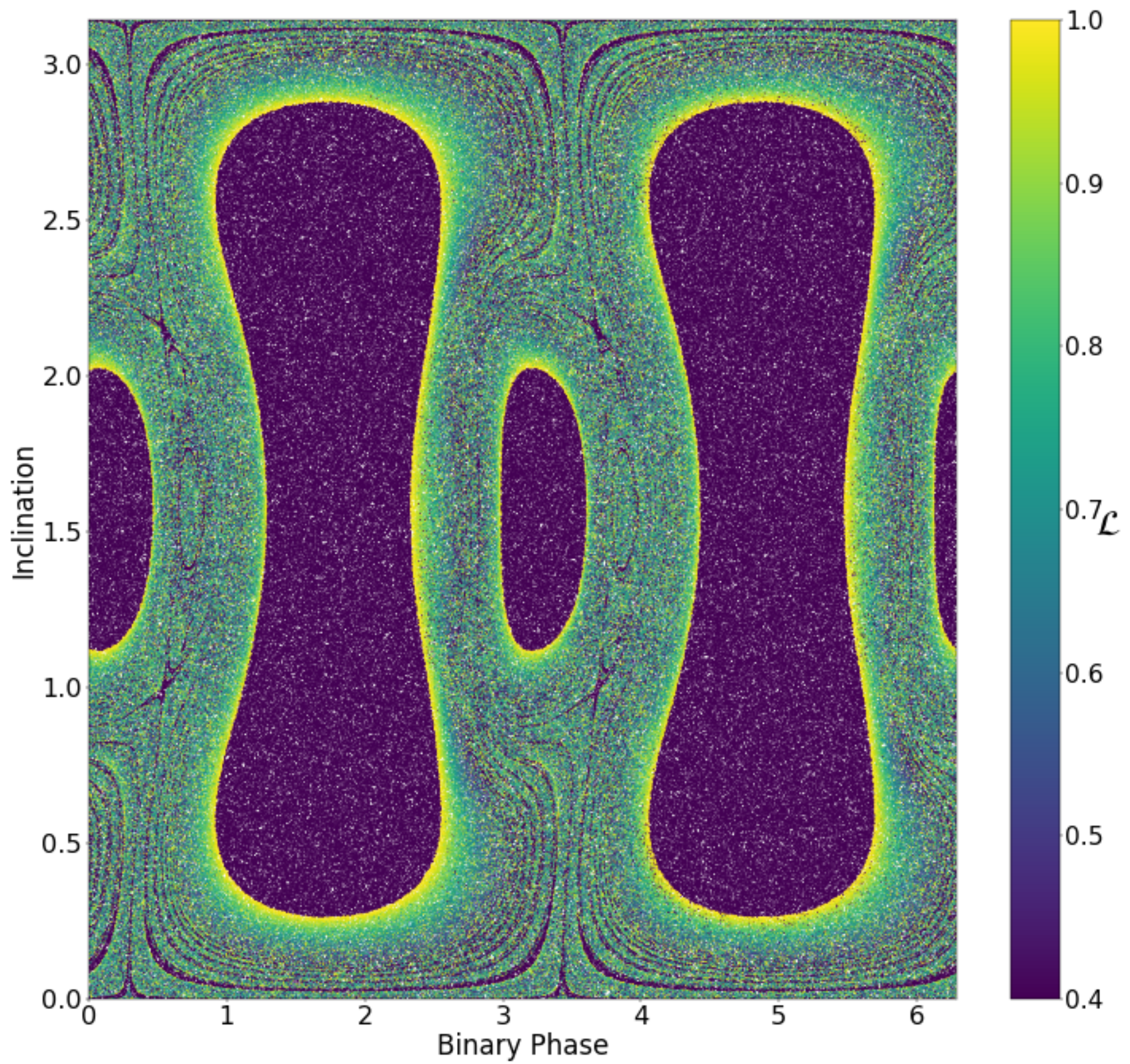}
            \caption[]%
            {{\small Phase space map color-coded according to the value of LF index $\mathcal{L}$ in the 15,15,15$M\textsubscript{\(\odot\)}$ system.}}    
            \label{fig:levy_equal_phase}
        \end{subfigure}
\caption{Three-body phase space maps color coded according to Scramble Density $\mathcal{S}$ and LF index $\mathcal{L}$.}
\end{figure*}

We also incorporate the calculation of the total number of scrambles in the TSUNAMI code. Based on our earlier discussion, we know that a scramble is defined as a peak above $R_{\mathcal{H}} = 0.33$. Therefore, the TSUNAMI code calculates the total number of scrambles by calculating the total number of peaks above $R_{\mathcal{H}} = 0.33$ in the homology radius data that is calculated from the position (x,y,z) data for the 3 masses at each time-step. Furthermore, we consider an excursion only if its duration lasts longer than $3\tau_{cr}$. We verified that our results converge at the excursion length $3\tau_{cr}$ and choosing smaller excursion lengths does not influence the results.

\subsection{Power-Law and Exponential Curve Fitting}\label{sec:model_fitting}

A reverse accumulation distribution $f(x)$ is a distribution where $f(x)$ is the fraction of values that exceed $x$. A general power-law reverse accumulation distribution is a two-parameter function given by, 
\begin{ceqn}
\begin{align}
   f(\tau_{D}) = c \cdot \tau_{D}^{-\alpha}
\end{align}
\end{ceqn}
It is easier to obtain the power-law index $\alpha$ by performing a linear fit in log-log space. Therefore, in this paper, to obtain $\alpha$, we perform a two-parametric linear fit to the cumulative lifetime distribution using the function,
\begin{ceqn}
\begin{align}
   \log_{10}f(\tau_{D}) = -\alpha \cdot \log_{10}\tau_{D} + \log_{10} c
\end{align}
\end{ceqn}

It is important to note that we are using reverse accumulation distribution instead of power-law integral distributions, where the general function is given by $f(x) = 1 - c\cdot x^{-\alpha}$. This is because $f(x) = c\cdot x^{-\alpha}$ (reverse accumulation) is linear in log-log space unlike the power-law integral distribution which is not linear in log-log space. However, in both cases, the power-law index $\alpha$ is the same. 

For the exponential distribution fitting, we also perform a two-parametric fitting using the function \citep[e.g.][]{valtonen06,leigh16,ibragimov18}:

\begin{ceqn}
\begin{align}
   f(\tau_{D}) = c \cdot e^{-\tau_{D}/\tau_{0}}
\end{align}
\end{ceqn}
where $\tau_{0}$ is the decay parameter. The relation between the decay parameter $\tau_{0}$ and the half-life $\tau_{1/2}$ is given by 
\begin{ceqn}
\begin{align}
   \tau_{1/2} = \tau_{0} \ln{2}
\end{align}
\end{ceqn}

Furthermore, in Section~\ref{decompose}, we propose a model for the general three-body lifetime distribution. To perform all the model fittings to the lifetime distributions, we use Bayesian Markov Chain Monte Carlo (MCMC) method using the \texttt{emcee} package \citep{emcee}. We assume flat priors and Gaussian errors in constructing the likelihood function. We report the median parameter values as the best fit parameter along with the estimated $1\sigma$ uncertainties. To judge the goodness of our respective fits, we use the reduced chi square statistic ($\chi_{\nu}^2$). A reduced chi square $\chi_{\nu}^2$ close to 1 is indicative of a good fit. 

In Section~\ref{decompose}, we compare different models for the general three-body lifetime distribution. Though, the reduced chi-square  $\chi^{2}_{\nu}$ is indicative of whether we have a good fit or not, to compare which model fits better, the Akaike Information Criterion (AIC) is a robust statistical criterion to judge how good a model is. This model was pioneered by \citet{akaike1974new} and is used extensively in many fields, including astrophysics, for model selection. The AIC is defined as 
\begin{ceqn}
\begin{align}
   \text{AIC} = -2 \ln{\mathcal{L}_{max}} + 2k  
\end{align}
\end{ceqn}
where $\mathcal{L}_{max}$ is the maximum likelihood achievable by the model under consideration and $k$ is the number of parameters in the model \citep{akaike1974new}. The model which minimizes the value of AIC is the best model. More details about AIC can be found in \citet{akaike1974new}, \citet{liddle07} and \citet{takeuchi2000}.

\section{Results} \label{results}

In this section, we apply our new metrics, namely the Scramble Density and LF index, to our simulations to study the phase space, half-lives and lifetime distribution tails of our three-body systems. In Section~\ref{sssec:levy_index_use} and  Section~\ref{sssec:scram_results}, we will apply the LF index and Scramble Density respectively to our sets of simulations and study their efficacy. In Section~\ref{sssec:chaos_tails}, using the set of ergodic interactions we isolate using Scramble Density, we discuss three-body lifetime distributions in the ergodic limit and fit exponential models to obtain respective half-lives $\tau_{1/2}$. Furthermore, in Section~\ref{sssec:levy_tail}, we study the power-law tails of the cumulative three-body lifetime distribution.

\begin{figure}
\includegraphics[width=\columnwidth]{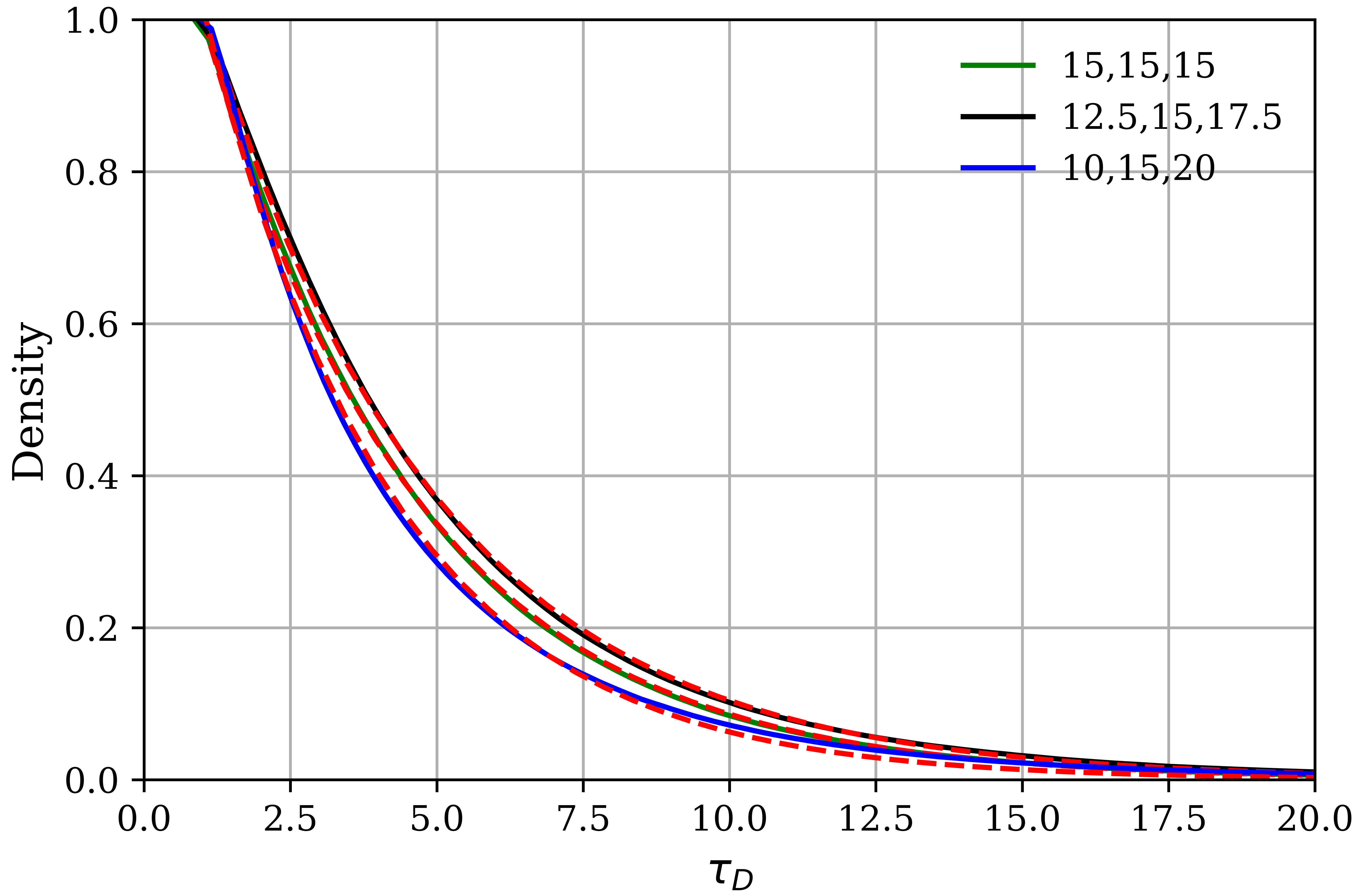}
\caption{Cumulative Lifetime Distributions (reverse accumulation) for the three systems in the ergodic limit. The red-dashed lines denote the exponential distribution fits with characteristic half-lives $\tau_{1/2}$ tabulated in Table~\ref{table:all_half_life}}
\label{fig:exp_tails_all}
\end{figure}

\begin{table}
\caption{Half-lives $\tau_{1/2}$ in units of $\tau\textsubscript{cr}$ and corresponding $\chi_{\nu}^2$ values for lifetime distribution for ergodic interactions}
\label{table:all_half_life}
\centering
\begin{tabular}{ccc}
\hline
 Masses($M\textsubscript{\(\odot\)}$) & $\tau_{1/2}$ & $\chi_{\nu}^2$ \\
\hline
15,15,15 & $2.580$ $\pm 0.011$ & $1.006$ \\
12.5,15,17.5 & $2.767$ $\pm 0.014$ & $1.001$ \\
10,15,20 & $2.261$ $\pm 0.011$ & $1.004$ \\
\hline
\end{tabular}
\end{table}

Before proceeding, it will be useful to give a brief prior on phase spaces in three-body systems. A phase space is a space that represents all the possible states of the system. In this case, we are considering the initial conditions phase space map. Therefore, each point in phase space represents a unique initial state of the system. Similar phase space maps in the context of three-body interactions were also proposed by \citet{hut83top}.

In our three-body simulations, each unique system is defined by a unique combination of inclination and binary phase for the initial configuration.  These two parameters range between $[0,\pi]$ and $[0,2\pi]$ respectively. Color coding the phase map according to a number of different parameters, including the disruption time ($\tau_{D}$) or the LF- and/or Scramble Density metrics, is insightful in understanding the fractal structures present in the phase space of the three-body system. For example, Figure~\ref{fig:fig3} is the phase space map for the 15,15,15$M\textsubscript{\(\odot\)}$ system color-coded according to the disruption time $\tau_{D}$ in units of $\tau\textsubscript{cr}$. Based on the coloring, one can see the black-colored islands of prompt interactions interspersed in the sea of chaos in phase space. We refer to the interactions in these regular islands of prompt interactions as ``ordered'' interactions (as opposed to chaotic). The smaller ordered islands can be seen in zoomed in versions of this phase space map, as in Figure~\ref{fig:zoom_small} or Figure~\ref{fig:zoom_equal}. We will discuss more relevant features of the phase space in the following sections, as they come up. 


\begin{table}
\caption{Half-lives $\tau_{1/2}$ in units of $\tau\textsubscript{cr}$ and corresponding $\chi_{\nu}^2$ values for the subset of lifetime distributions corresponding to ergodic interactions in which the same particle mass is ejected. In the Ejection Mass column, for the equal mass system, $15 (0)$ denotes the mass that was initially a single particle, whereas $15(1)$ and $15(2)$ denote the particles that initially composed the binary}.
\label{table:half_life_eject}
\centering
\begin{tabular}{cccc}
\hline
 Masses($M\textsubscript{\(\odot\)}$) & Ejection Mass($M\textsubscript{\(\odot\)}$) & $\tau_{1/2}$ & $\chi_{\nu}^2$  \\
\hline
& 15 (0) & $2.585\pm 0.018$ & 1.007  \\
15,15,15 & 15 (1) & $2.572\pm0.010$ & 1.006  \\
 & 15 (2) & $2.578\pm0.013$ & 0.996 \\
 & 12.5 & $2.764\pm0.015$ & 1.006 \\
12.5,15,17.5 & 15 & $2.771\pm0.015$ & 1.001  \\
 & 17.5 & $2.743\pm 0.011$ & 0.999 \\
 & 10 & $2.267\pm0.013$ & 1.004 \\
10,15,20 & 15 & $2.291\pm 0.014$ & 1.004 \\
 & 20 & $2.206\pm 0.014$ & 0.994  \\
\hline
\end{tabular}
\end{table}

\subsection{Efficacy of the LF Index}
\label{sssec:levy_index_use}

In Figure~\ref{fig:fig3}, one observes how the boundary of the ordered islands are populated by long lived interactions (yellow/white rims according to color-map). As we know that the lifetime distribution tail is composed mainly of L\'evy flights, we predict that these long-lived interactions on the ``chaos-order'' border are L\'evy flight interactions. Using the LF index, we can verify whether these border interactions correspond to the L\'evy flights. Figure~\ref{fig:levy_equal_phase} shows the same phase space map color-coded according to the LF index value. One indeed observes that the interactions corresponding to a high LF index value (close to 1) populate the ``chaos-order'' border in phase space, implying that they are indeed L\'evy flight interactions.

\subsection{Scramble Density}
\label{sssec:scram_results}

The dark-colored islands in Figure~\ref{fig:fig3} that represent the prompt interactions are the ordered interactions. The rest of the interactions in phase space that do not conglomerate into any sort of structure should represent the ergodic interactions. This is also evident from the very definition of ergodicity; that is, sensitivity to initial conditions and the ability to forget initial conditions over the course of the interaction. Of particular interest to us is the first feature, that is, sensitivity to initial conditions. In the ordered regions, we should observe that slight differences in the initial conditions should not lead to drastically different outcomes.  This is exactly what we see in the dark-colored islands. However, in the rest of the phase space, we see that even when we change the initial conditions slightly, the disruption time changes drastically in a random fashion as denoted by the color. This change in disruption time directly translates to a completely different interaction outcome signifying the sensitivity to initial conditions. Therefore, if the Scramble Density $\mathcal{S}$ is representative of the ergodicity of the system, then we should be able to identify those ergodic areas of the phase space by values of the $\mathcal{S}$ as discussed in Section~\ref{sssec:scram}. 

In Figure~\ref{fig:scram_index_phase} for the equal mass system, we observe\footnote{The color scales have been adjusted such that it is easier to see this effect.} that interactions with $\mathcal{S} > 10.64$ occupy the regions of chaos we earlier highlighted in the phase space map. There is no formal method in which we obtained this number. We observed a range of cutoff values and the corresponding regions in the phase space map and reported the value most accurately describing the ergodic regions. Though the similar plots for the other mass ratios are not shown here, we have tabulated the values of the Scramble Density cutoff for the ergodic subset in Table~\ref{table:dense_values}. It is interesting to note is that when using $\tau_{D}$ in units of years while calculating the Scramble Density (instead of units of $\tau_{cr}$), the same cutoff value of 0.22 $\text{yr}^{-1}$ is observed for all the 3 systems under consideration. 

\begin{table}
\caption{Scramble Density $\mathcal{S}$ cutoffs for the ergodic subset for different systems.}
\label{table:dense_values}
\centering
\begin{tabular}{cc}
\hline
 Masses($M\textsubscript{\(\odot\)}$) & $\mathcal{S}$ \\
\hline
15,15,15 & 10.64 \\
12.5,15,17.5 & 8.94  \\
10,15,20 & 7.70 \\
\hline
\end{tabular}
\end{table}

\subsection{Cumulative Lifetime Distributions in the Ergodic limit}
\label{sssec:chaos_tails}

As discussed in the previous section, we observe that interactions with Scramble Density $\mathcal{S}$ above the cutoff values shown in Table~\ref{table:dense_values} occupy the ergodic regions of phase space. We will refer to these ergodic interactions as `EC-type interactions'. The lifetimes of EC-type interactions follow an exponential curve distribution like radioactive decay as seen in  Figure~\ref{fig:exp_tails_all}. 

Therefore, in the ergodic limit, the three-body lifetime distribution follows an exponential model analogous to the radioactive decay model. Fitting exponential distribution models to these cumulative lifetime distributions, we obtain the half-lives $\tau_{1/2}$ as shown in Table~\ref{table:all_half_life}. Therefore, even though \citet{valtonen88} supposed that the lifetime distribution of three-body is an exponential distribution like radioactive decay or \citet{mikkola07} fit exponential models to lifetime distributions of equal mass systems, the exponential distribution is only fully realized in the ergodic limit. Considering it to be exponential in the non-ergodic limit is incorrect due to the L\'evy flight interactions that follow a power-law/algebraic distribution. We will discuss this in more detail in Section~\ref{sssec:levy_tail}.

Another interesting observation is that the lifetime distributions for different ejection types in the ergodic limit also follow exponential distributions. The ejection type under discussion here is based on which of the 3 masses in the system is being ejected. Therefore, there are 3 ejection types for each system. Table~\ref{table:half_life_eject} contains the half-lives $\tau_{1/2}$ in $\tau\textsubscript{cr}$ for the 3 ejection types for each three-body system under consideration. One observes that in the ergodic limit, the half-lives $\tau_{1/2}$ for different ejection types in system are within $1,2\sigma$ of each other as one can see in Table~\ref{table:half_life_eject}, except for the ejection of 20$M\textsubscript{\(\odot\)}$ in the 10,15,20$M\textsubscript{\(\odot\)}$ system even though they are qualitatively very similar. This is indicative of the question of the validity of Scramble Density and ergodic assumptions in the unequal mass limit, which will be discussed in more depth in Section~\ref{erg_cut} and Section~\ref{sssec:half_ratio}. In any case, we do observe that the ratio of the half-lives $\tau_{1/2}$ between different ejection types in the ergodic limit is essentially 1:1:1.

\begin{table}
\caption{Power law indices $\alpha$ for reverse accumulation lifetime distribution tails ($\tau_{D} > 100$) of all interactions with the corresponding $\chi_{\nu}^2$ values for the fit.}
\label{table:power_entire}
\centering
\begin{tabular}{ccc}
\hline
 Masses($M\textsubscript{\(\odot\)}$) & $\alpha$ & $\chi_{\nu}^2$ \\
\hline
15,15,15 & $0.783$ $\pm 0.005$ & 1.013 \\
12.5,15,17.5 & $0.764$ $\pm 0.008$ & 1.019 \\
10,15,20 & $0.790$ $\pm 0.006$ & 1.034 \\
\hline
\end{tabular}
\end{table}

\begin{figure}
\includegraphics[width=\columnwidth]{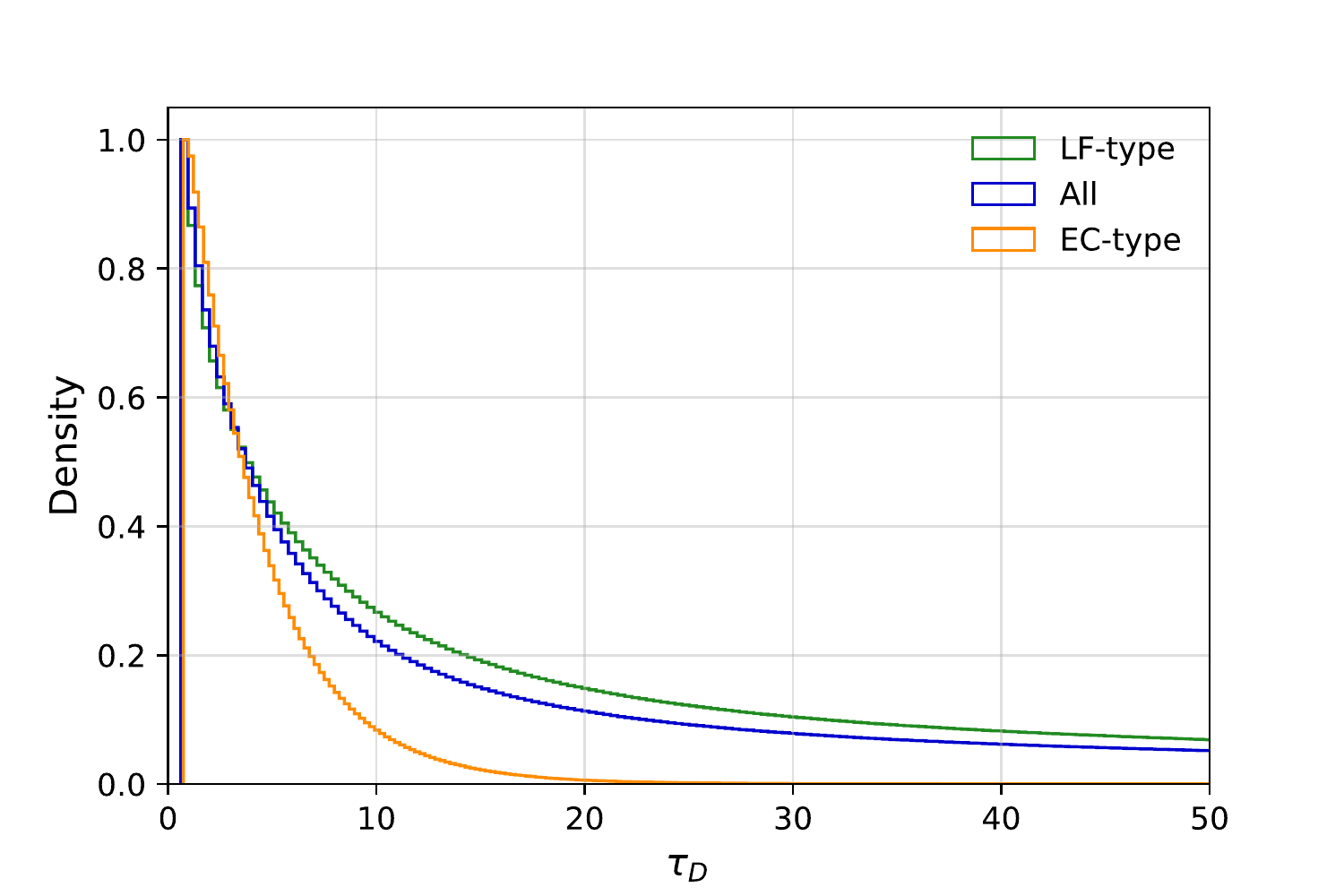}
\caption{A comparison plot showing the reverse accumulation distributions plots for all interactions, L\'evy flight and EC-type interactions respectively for 15,15,15$M\textsubscript{\(\odot\)}$ system}
\label{fig:compare_nolog}
\end{figure}

\begin{figure}
\includegraphics[width=\columnwidth]{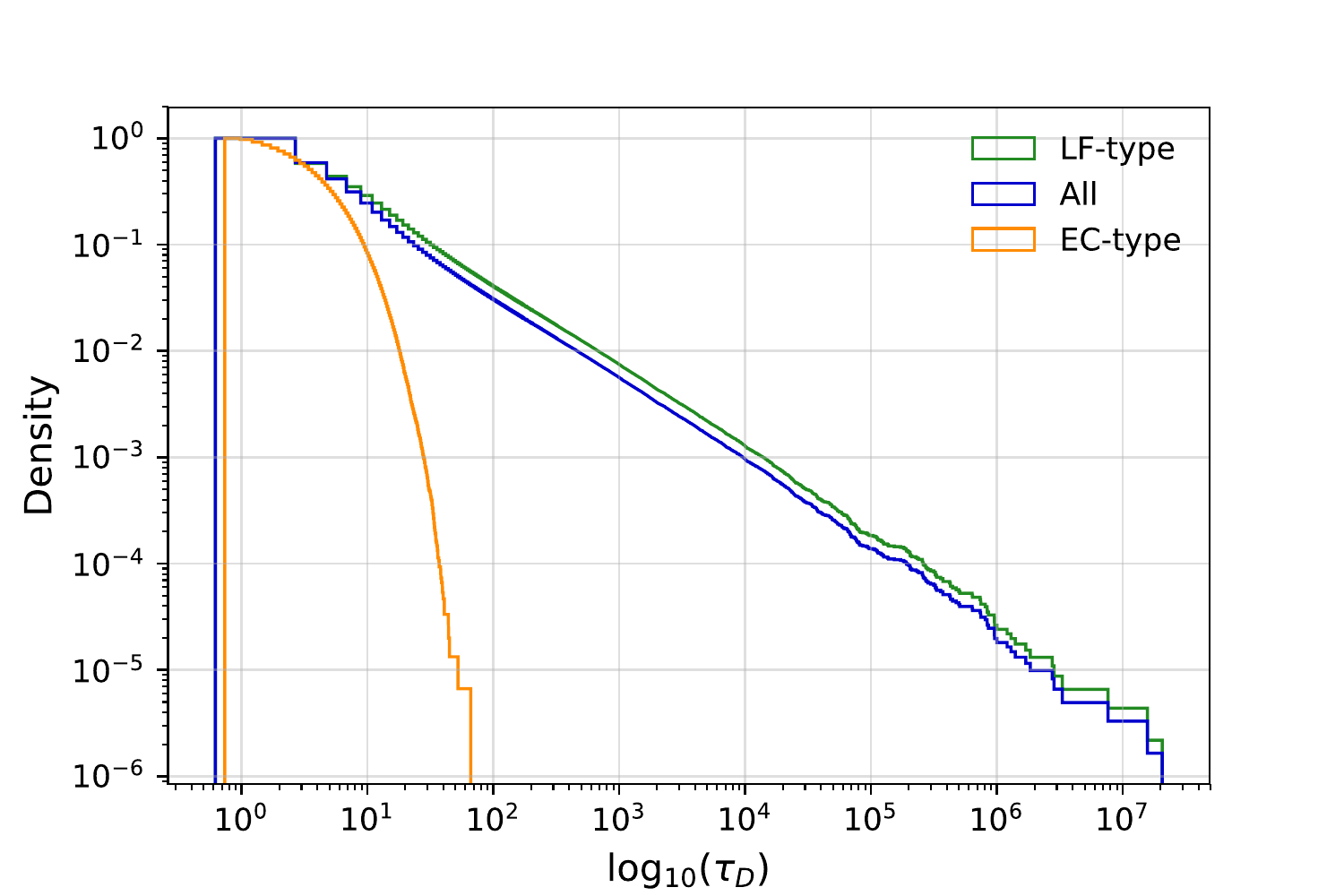}
\caption{A comparison plot showing the reverse accumulation distributions plots for all interactions, L\'evy flight and EC-type interactions respectively for 15,15,15$M\textsubscript{\(\odot\)}$ system in log-log scale.}
\label{fig:compare_log}
\end{figure}

\subsection{Power-Law analysis of Cumulative Lifetime Distribution Tails}
\label{sssec:levy_tail}


As discussed in \citet{orlov10} and \citet{shevchenko10}, the lifetime distribution tail of three-body interactions has a heavy-tail distribution with the initial part of the distribution being exponential. This phenomena is because of L\'evy flight interactions as discussed earlier in Section~\ref{sssec:levy}. We have already see one part, that is the exponential decay distribution of the EC-type interactions, unfold in Section~\ref{sssec:chaos_tails}. Now, we will study the later part of the distribution, namely the tail, that is composed L\'evy flight interactions.

\begin{figure*}
        \centering
        \begin{subfigure}[b]{0.485\textwidth}
            \centering
            \includegraphics[width=\textwidth]{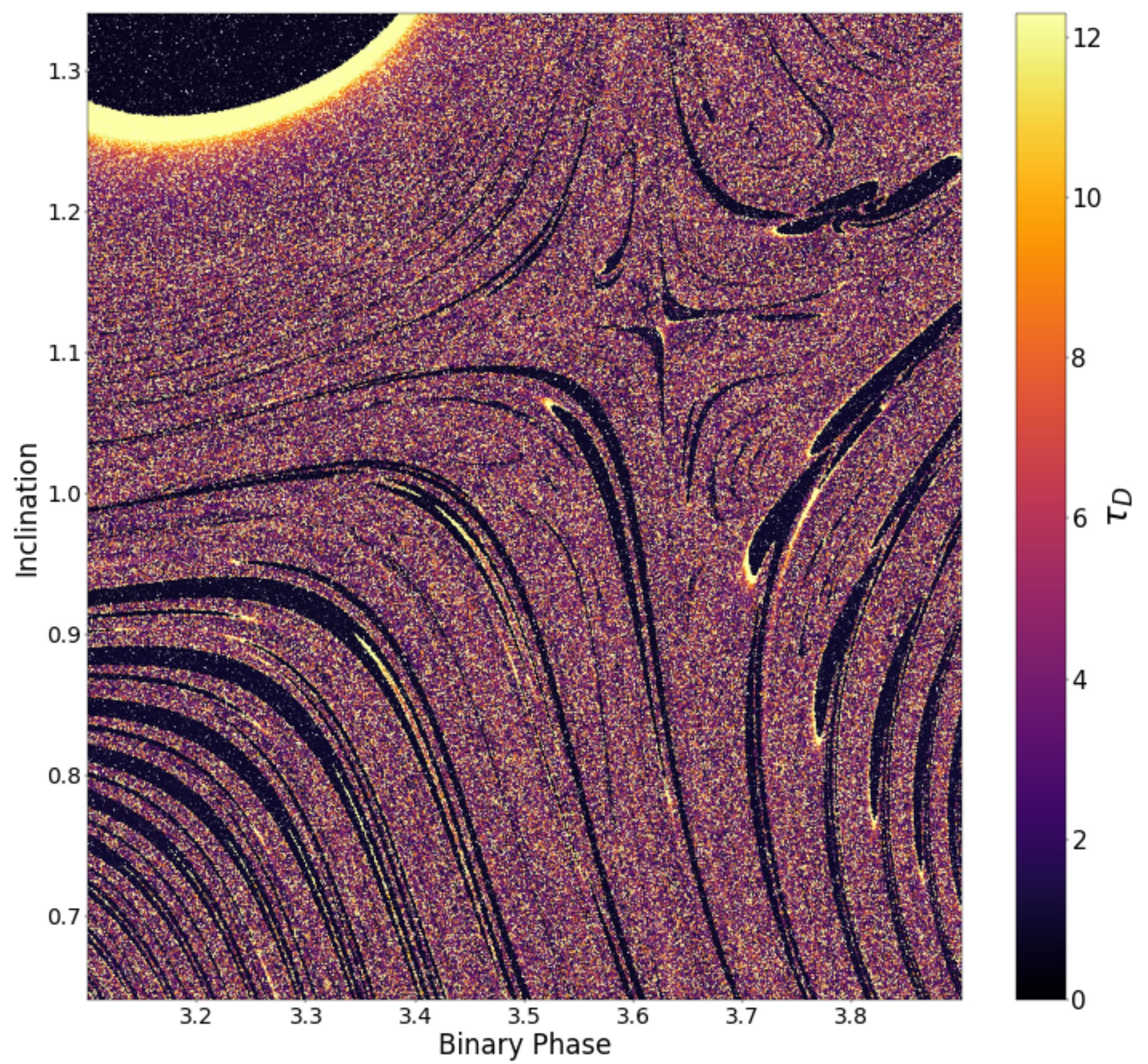}
            \caption[]%
            {{\small Zoomed-in phase space region for 12.5,15,17.5 $M\textsubscript{\(\odot\)}$ system color-coded according to $\tau_{D}$.}}    
            \label{fig:zoom_small}
        \end{subfigure}
        \hfill
        \begin{subfigure}[b]{0.485\textwidth}  
            \centering 
            \includegraphics[width=\textwidth]{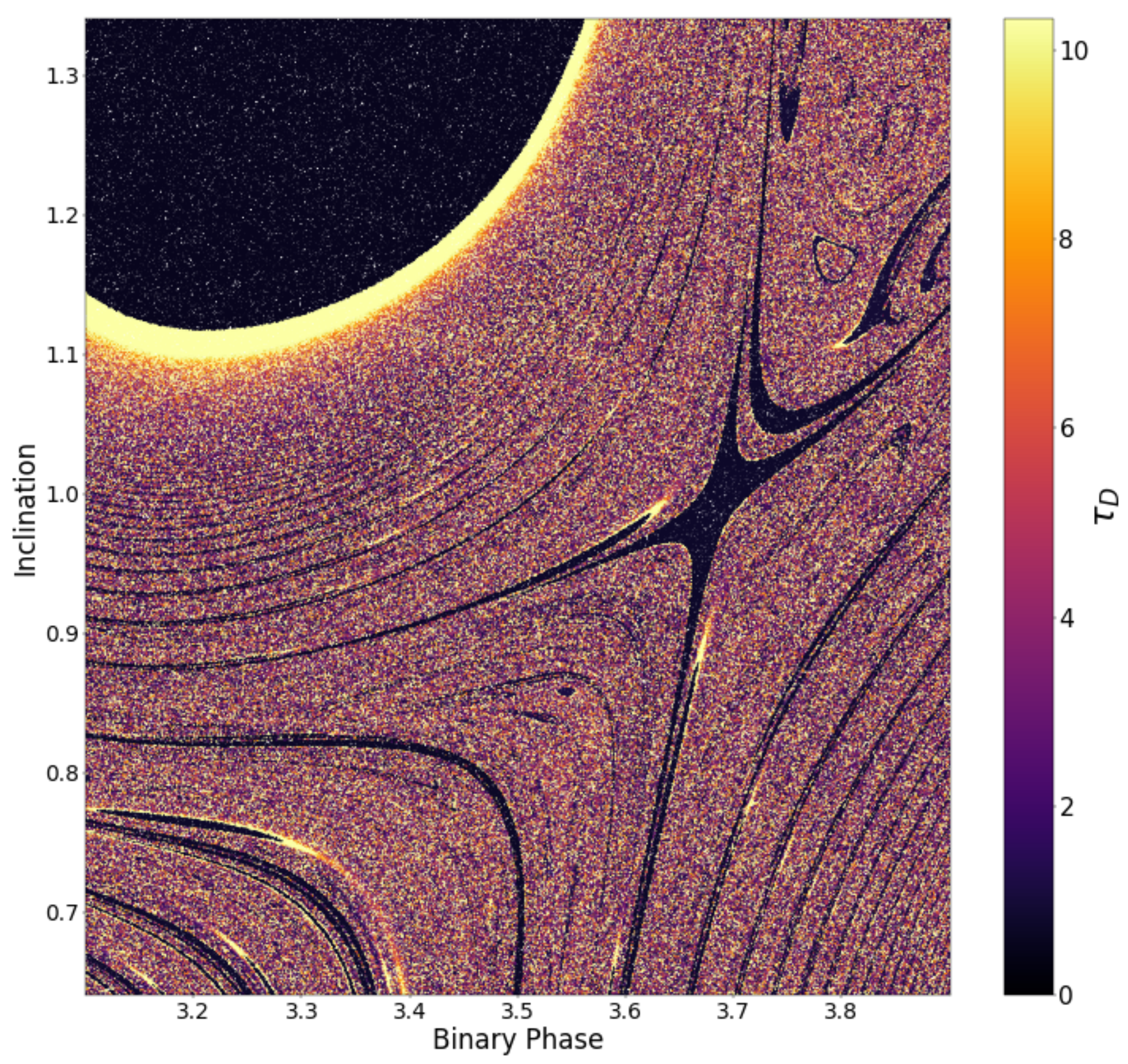}
            \caption[]%
            {{\small Zoomed-in phase space region for 15,15 and 15$M\textsubscript{\(\odot\)}$ system color-coded according to $\tau_{D}$.}}    
            \label{fig:zoom_equal}
        \end{subfigure}
        \vskip\baselineskip
        \begin{subfigure}[b]{0.485\textwidth}   
            \centering 
            \includegraphics[width=\textwidth]{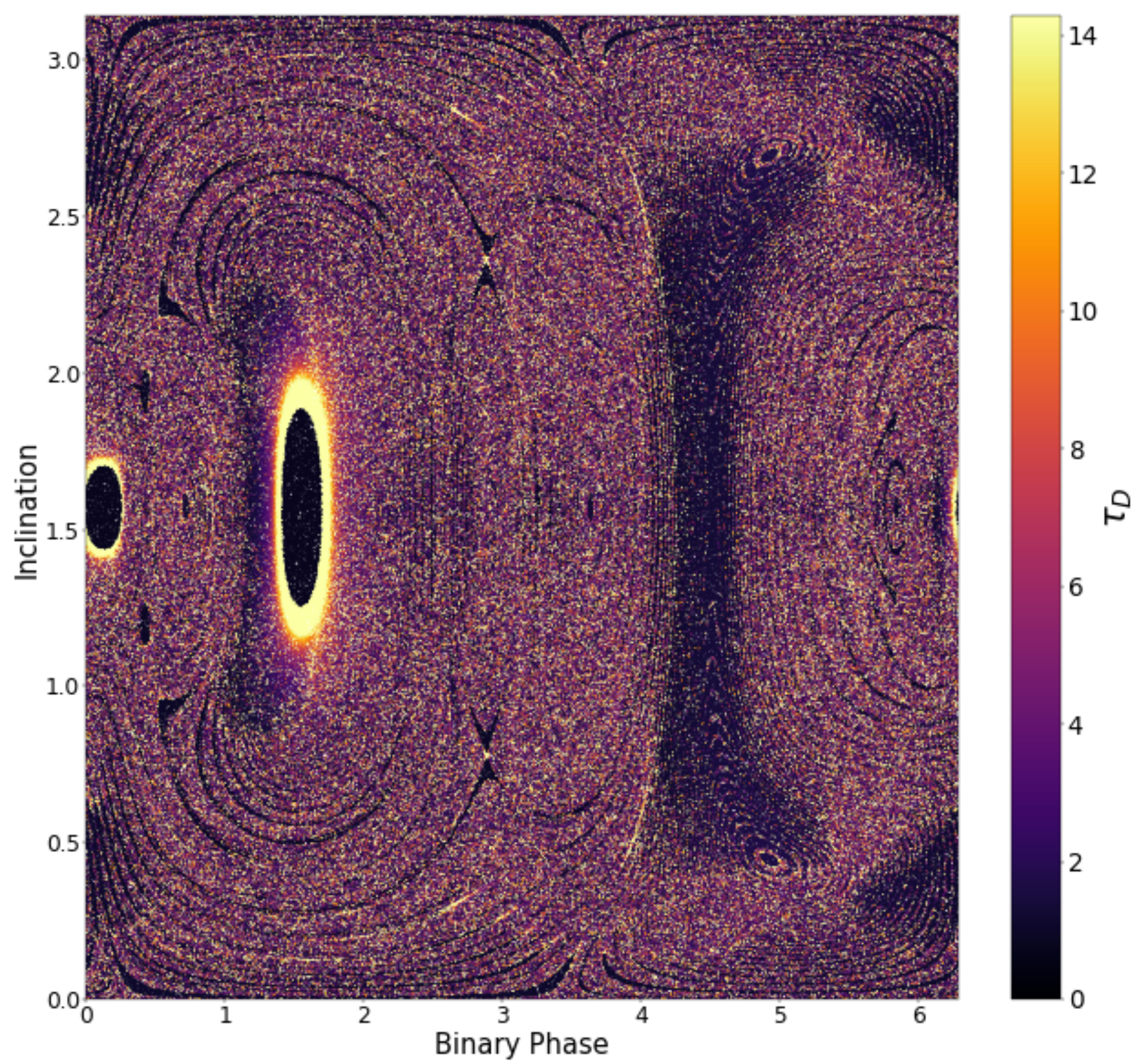}
            \caption[]%
            {{\small Phase space map for 10,15 and 20$M\textsubscript{\(\odot\)}$ system color-coded according to $\tau_{D}$.}}    
            \label{fig:global_broken}
        \end{subfigure}
        \hfill
        \begin{subfigure}[b]{0.485\textwidth}   
            \centering 
            \includegraphics[width=\textwidth]{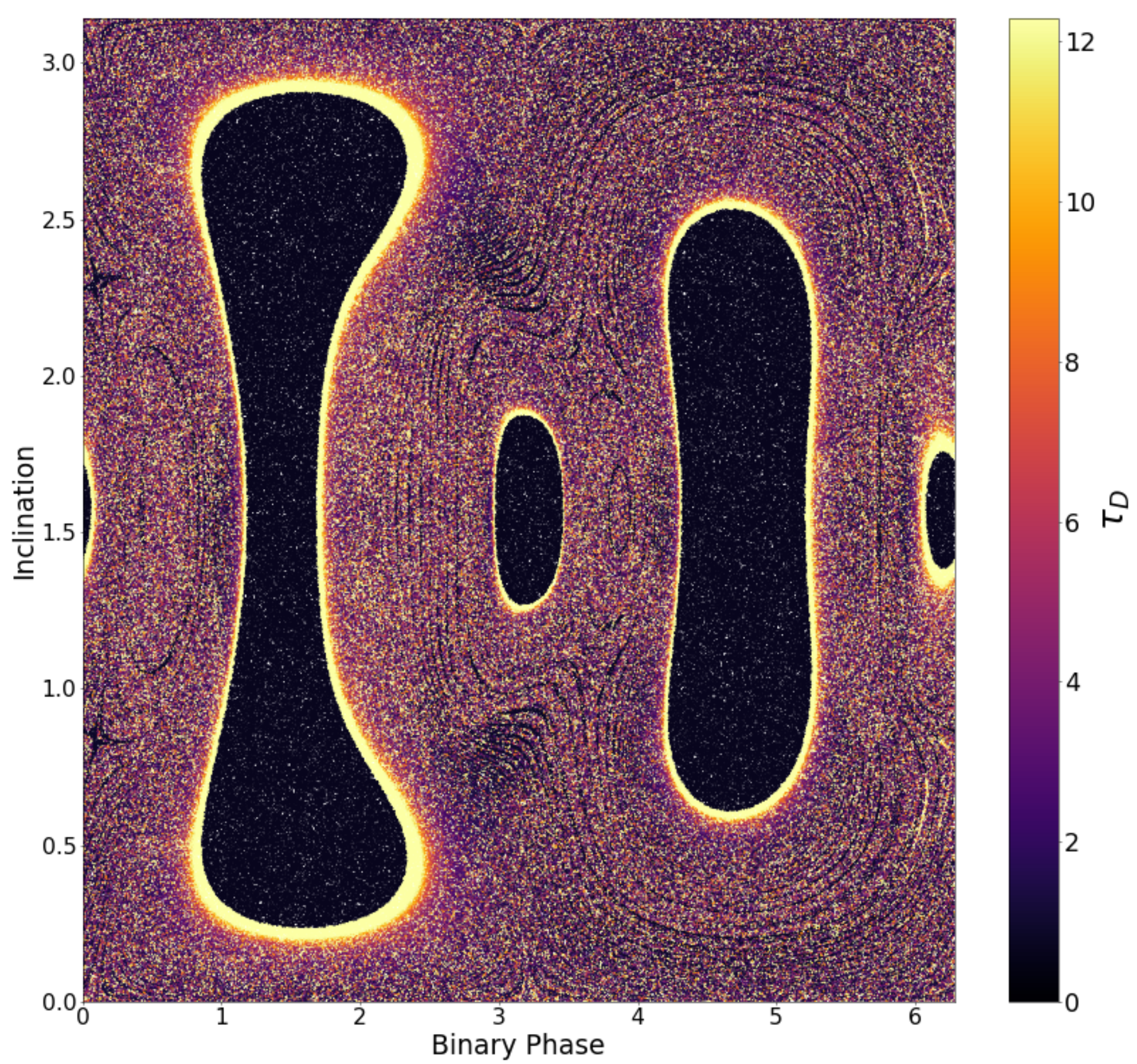}
            \caption[]%
            {{\small Phase space map for 12.5,15 and 17.5$M\textsubscript{\(\odot\)}$ system color-coded according to $\tau_{D}$.}}    
            \label{fig:global_small}
        \end{subfigure}
\caption{Three-body phase space maps color-coded according to disruption time $\tau_{D}$ for 12.5,15,17.5$M\textsubscript{\(\odot\)}$ and 10,15,20$M\textsubscript{\(\odot\)}$ system.}
\end{figure*}

Unlike, EC-type interactions that follow an exponential distribution, L\'evy flights follow pure power-law distributions. As mentioned earlier in Section~\ref{sec:model_fitting}, we perform the power-law fitting in log-log scale as we obtain a linear relation which is easier to fit. We perform the tail power law fitting over the range $10^2 < \tau_{D} < 10^5$. Attempting to fit the end of the distribution tail ($\tau_{D} > 10^5$) isn't a reliable approach because there aren't enough interactions to populate the distribution properly.   

Figure~\ref{fig:compare_log} shows the lifetime distributions for all interactions, LF-type and EC-type interactions respectively for the 15,15,15$M\textsubscript{\(\odot\)}$ system in log-log space. We indeed see that the distribution tails are described very well by a power-laws as they appear linear in log-log space. Table~\ref{table:power_entire} shows the fitted power-law indices for the cumulative lifetime distribution tails. What this power-law index implies in the context of L\'evy flights will be discussed in detail in Section~\ref{levy_types}.
 
What is interesting to observe is that in the distribution for all the interactions in log-log space, it is hard to make out the initial small bump due to EC-type interactions. However, once the L\'evy flight interactions are removed, the underlying EC-type interaction distributions become prominent. This also highlights the dominant behavior of L\'evy flights in the three-body system.
\textit{We observe that removing LF-type interactions removes a major part of the distribution tail, providing evidence for the fact that L\'evy flights contribute non-negligibly to the heavy-tailed nature of the lifetime distribution}. We have therefore shown conclusive evidence for the fact that L\'evy flights indeed occupy the tail of the lifetime distribution and are indeed responsible for the heavy-tailed nature of the three-body problem.

\begin{figure*}
    \centering
        \begin{subfigure}[b]{0.48\textwidth}
            \centering
            \includegraphics[width=\textwidth]{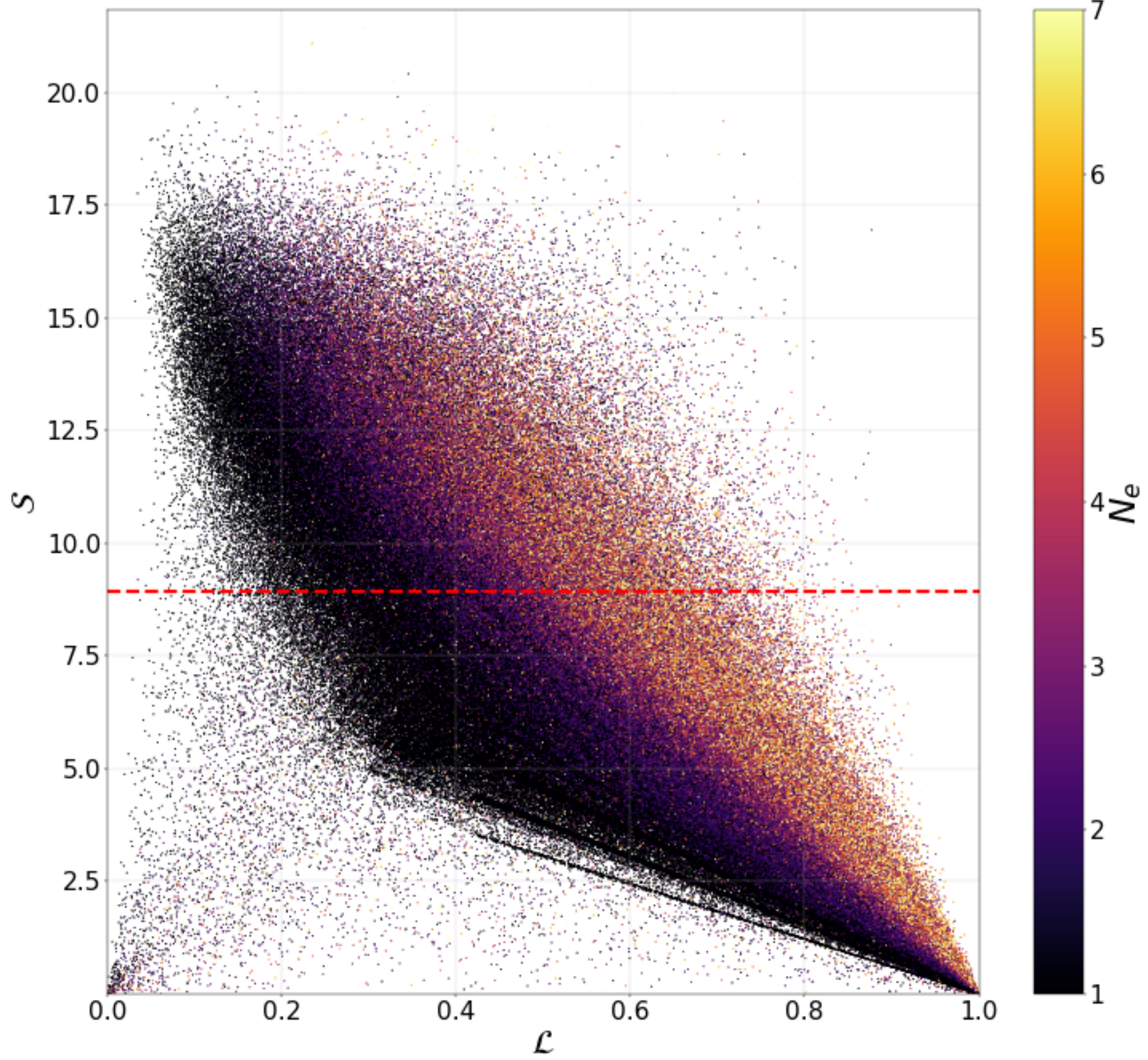}
            \caption[]%
            {{\small All three-body interactions (except ordered) in $\mathcal{S}$ vs. $\mathcal{L}$ space for 12.5,15,17.5 $M\textsubscript{\(\odot\)}$ system, color-coded according to the number of excursions $N_{e}$. The horizontal red-dashed line denotes the ergodic cutoff value for the Scramble Density $\mathcal{S}$.}}    
            \label{fig:ls_all}
        \end{subfigure}
        \hfill
        \begin{subfigure}[b]{0.48\textwidth}  
            \centering 
            \includegraphics[width=\textwidth]{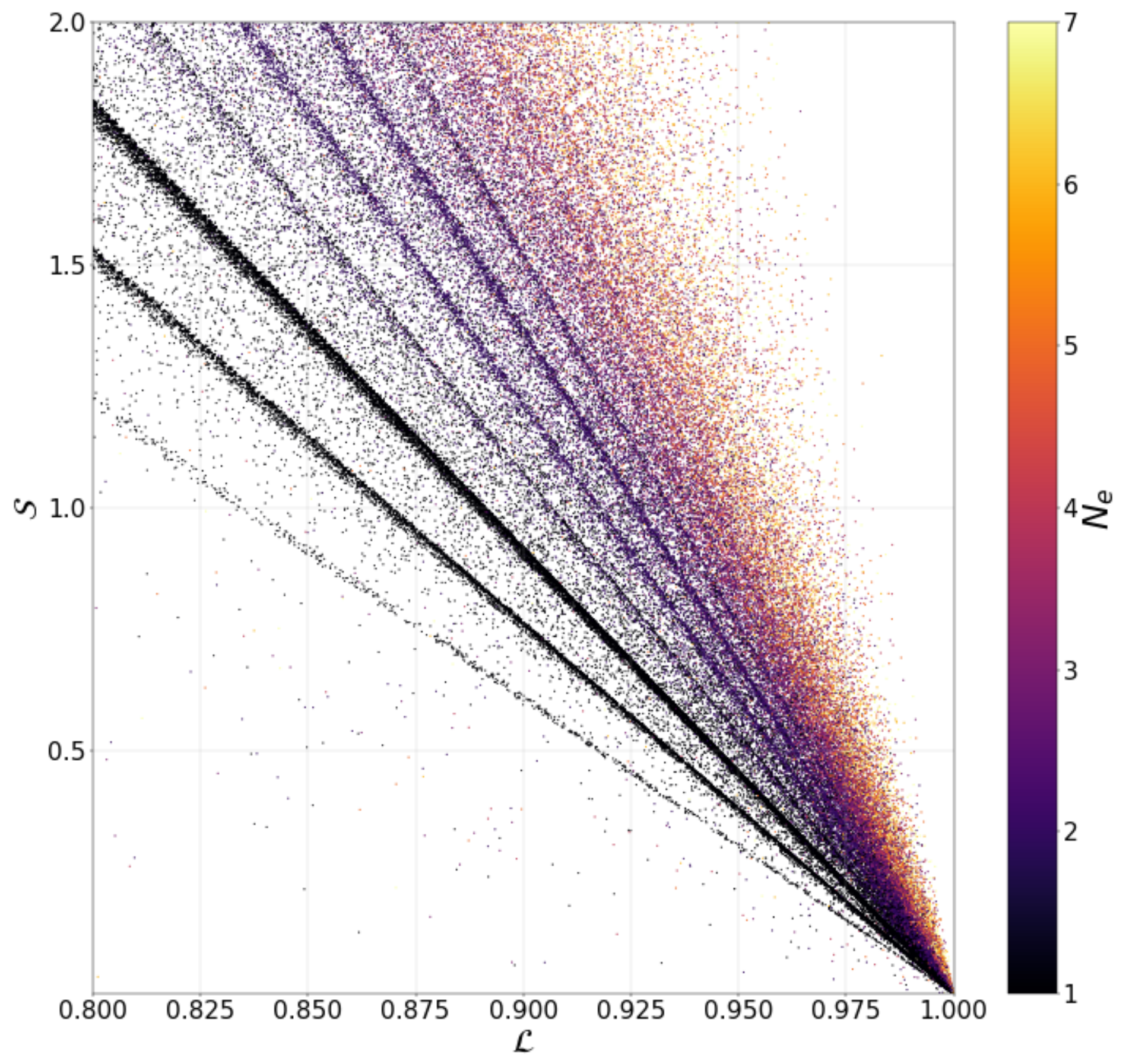}
            \caption[]%
            {{\small Zoomed-in version of $\mathcal{S}$ vs. $\mathcal{L}$ space for all three-body interactions (except ordered) for the 12.5,15,17.5 $M\textsubscript{\(\odot\)}$ system, color-coded according to the number of excursions $N_{e}$. 
            }}    
            \label{fig:ls_zoom}
        \end{subfigure}
\caption{Three-body interactions for 12.5,15,17.5$M\textsubscript{\(\odot\)}$ in $\mathcal{S}$ vs. $\mathcal{L}$ space.}
\end{figure*}

\begin{figure*}
\includegraphics[width=2\columnwidth]{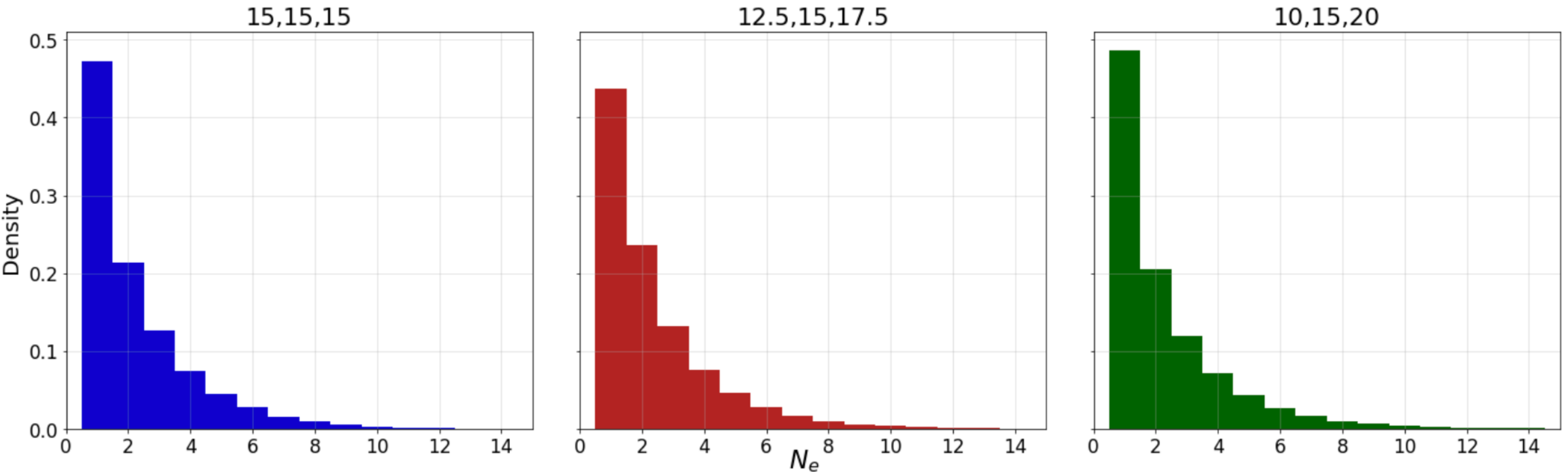}
\caption{Normalized histogram distributions for the number of excursions $N_{e}$ in L\'evy flights in the 3 mass combinations considered here.}
\label{fig:bins_all}
\end{figure*}


\section{Discussion} \label{discussion}

In this section, we discuss the significance of our results for chaos in the general three-body problem and the lifetime distributions. In Section~\ref{ls_rel}, we discuss the relation between Scramble Density and LF index. In Section~\ref{erg_cut}, we discuss the validity of Scramble Density as parameter for ergodicity in the unequal mass limit. In Section~\ref{levy_types}, we discuss the types of L\'evy flights in the general three-body problem in context of the analysis presented in \citet{shevchenko10}. In Section~\ref{tail}, we study the physical origin of the L\'evy flights and introduce the \textit{three-body relaxation} mechanism. In Section~\ref{decompose}, we propose a new model to study general three-body distributions to extract information about the EC-type and L\'evy flight interactions from the distribution. In Section~\ref{sssec:half_ratio}, we study the ratio of half-lives and the ejection probabilities in the ergodic limit.

\subsection{The Scramble Density ($\mathcal{S}$) - LF index ($\mathcal{L}$) relation}
\label{ls_rel}

Based on the definitions of Scramble Density and LF index, the relation between the two metrics is essentially the relation between $N_{T}$, the total number of scrambles, and $\tau_{H}$, the total amount of time spent in the hierarchical state. Therefore, one expects that the number of scrambles should be inversely proportional to the amount of time spent in the hierarchical state by the interaction. This is because scrambles can only occur when there is no hierarchical structure present, as discussed in Section~\ref{sssec:scram}. This inverse proportionality relation is exactly what we observe in Figure~\ref{fig:ls_all}, in the zoomed-in version in Figure~\ref{fig:ls_zoom}. These figures show a scatter plot of the Scramble Density and LF index values for the EC-type and L\'evy flight interactions for the 12.5,15,17.5 $M\textsubscript{\(\odot\)}$ system. Similar plots (not shown here) are observed for the 15,15,15 $M\textsubscript{\(\odot\)}$ and 10,15,20 $M\textsubscript{\(\odot\)}$ systems. The horizontal red-dashed line in Figure~\ref{fig:ls_all} is the Scramble Density ergodic cutoff value for the 12.5,15,17.5 $M\textsubscript{\(\odot\)}$ case. We will call the region below this line in the $\mathcal{S}$ vs. $\mathcal{L}$ space as the ``L\'evy flight regime'' and the region above this line as the ``ergodic regime''. 

There are a number of interesting features that we observe. First, we observe a dominant negative linear relation between the 2 metrics, which is expected. There is a certain width/thickness to this relation due to the intrinsic scatter in it. This intrinsic scatter in the interactions populating the $\mathcal{S}$ vs. $\mathcal{L}$ space is reflective of chaos in three-body processes. The linear relation is more pronounced in the L\'evy flight regime, as seen in Figure~\ref{fig:ls_zoom}. We see a number of linear relations with each line corresponding to the number of excursions in the L\'evy flight. This linear relation becomes less prominent as the number of excursions in the L\'evy flight increases. We do not yet know the physical origins of these relations, however, we hypothesize that these relations in the L\'evy flight regime have to do with the intricate fractal substructure present at the ``chaos-order'', as described by \citet{boris83}. In the ergodic regime, these linear relations no longer exist.

Second, the steepness of the linear relationship between Scramble Density and LF index increases as the number of excursions in an interaction increases as seen in both the figures. Therefore, interactions with the same LF index have a higher Scramble Density if they have more excursions over the course of their evolution. Physically, this makes sense because even though different interactions can spend the same duration in the hierarchical regime, more excursions imply that the single mass has returned more times to the binary pair, resulting in a higher number of scrambles and hence a higher Scramble Density. Therefore, interactions with larger numbers of excursions have a steeper relation in Scramble Density vs. LF index space.  

Third, in the L\'evy flight regime shown in Figure~\ref{fig:ls_all}, we observe 2 kinds of interactions. The first kind is indicated by the conglomerates along the linear relation discussed earlier. This is where a majority of the L\'evy flight interactions lie. However, there is another small ensemble of L\'evy flight interactions that originate on the opposite spectrum of the LF index and seem to be positively correlated with Scramble Density, unlike the dominant kind. This could provide a natural framework to distinguish between the 2 kinds of L\'evy flights in a general three-body system. We will discuss this in more detail in Section~\ref{levy_types}.

\subsection{Ergodic Cutoff in the Unequal Mass limit}
\label{erg_cut}

\begin{table}
\caption{Average number of scrambles in an interaction. }
\label{table:mean_scram}
\centering
\begin{tabular}{cc}
\hline
 Masses($M\textsubscript{\(\odot\)}$) & $N_{T}$ \\
\hline
15,15,15 & 37.8 \\
12.5,15,17.5 & 36.6  \\
10,15,20 & 35.1 \\
\hline
\end{tabular}
\end{table}

The value of the Scramble Density index gives an indication of the total number of scrambles in an interaction. We hypothesize that in the unequal mass limit, the scramble density becomes a poorer proxy for chaos. This is because in systems with very unequal distributions of masses, it is difficult for the lightest mass to disrupt the two heavier masses from being a pairwise binary. Only weak perturbations of the binary pair will occur and hence no scrambles. An example of this is the Sun-Jupiter-comet three-body system. Therefore, as the distribution of masses gets more unequal, one would expect the number of scrambles to decrease as well. We see this effect to some extent in Table~\ref{table:mean_scram} where we see that the average number of scrambles in an interaction decreases as we go towards the unequal mass systems. 

This suggests that different heuristic measures for chaos will be necessary in systems that are far from the equal-mass limit. Further investigation that is beyond the scope of the current paper will be needed to fully test the efficacy and robustness of the Scramble Density parameter in the unequal mass limit.

\subsection{L\'evy flights in the general three-body problem}
\label{levy_types}

As discussed in Section~\ref{ls_rel}, we observed 2 kinds of interactions in the L\'evy flight regime. Could these 2 kinds of interactions be the kinds of L\'evy flights we were looking for in the general three-body system? Can the second, less-dominant kind of interaction even be considered as a L\'evy flight? Can we extrapolate \citet{shevchenko10}\textquotesingle s LF1-type and LF2-type L\'evy flight types in Keplerian map dynamics to the general three-body scenario?

To begin answering these questions, let us reconsider the LF1-type and LF2-type interactions (discussed earlier in Section~\ref{sssec:levy}). Figure~\ref{fig:bins_all} shows the distribution for the number of excursions, $N_{e}$, in L\'evy flights in the 3 systems under consideration. We observe that a dominant portion of L\'evy flights have very few excursions ($\approx$ 1 - 3) during their lifetime, consequently corresponding to LF1-type interactions. Therefore, LF1-type interactions that exist in Keplerian map dynamics also exist in the general three-body scenario. However, unlike LF1-type, we do not observe LF2-type interactions. To get a physical sense of what LF2-type interactions look like in a Kepler map, we refer to Figure 1 in \citet{shevchenko10}. Their Figure 1 shows a system with both LF1-type and LF2-type interactions. The LF2-type motion appears as a series of $\approx$300 excursions of similar orbital size and period. In the general three-body system in our scenario, we do not see as many excursions in our interactions, as seen in Figure~\ref{fig:bins_all}, let alone in consecutive order. Therefore, we can extrapolate LF1-type interactions to the general three-body problem, but not LF2-type interactions.  Subsequent excursions tend to be more separated in orbital size and period for systems with particles of comparable mass (assuming no strong hierarchies are imposed).

It also physically makes sense why LF2-type interactions cannot be found in general three-body systems like ours. As discussed earlier in Section~\ref{erg_cut}, in the unequal mass limit, only weak perturbations in the heavier mass binary pair are possible by the lighter mass and hence there is an increasing tendency for excursions instead of scrambles. Therefore, in such a limit, one will observe a large number of excursions instead of scrambles, which would have otherwise disrupted the series of excursions. When there is a more democratic distribution of masses, as in the 3 systems considered in this paper, it is possible to have strong perturbations in the binary pair by the single mass and the masses are closer to equipartition, resulting in more scrambles. Therefore, in place of LF2-type behavior, one observes either scrambles or a \textit{three-body relaxation process} (described in more detail in Section~\ref{tail}).

What do the power-law indices of our lifetime distributions tell us? Based on Table~\ref{table:power_entire}, we have a power-law index in the range $\approx 0.76-0.79$ \footnote{The distribution has a negative power-law index but, as we are defining the index as $x^{-\alpha}$, we report positive values of $\alpha$} for our systems. \citet{shevchenko10} predicted that the power-law tails in LF1-dominated systems is $\approx 2/3$. In our systems, we do indeed see a power-law index close to $2/3$. Power-law indices similar to $2/3$ were also observed in other similar studies regarding tails of lifetime distributions in the general three-body problem. For instance, in \citet{orlov10}, equal mass three-body systems with randomized conditions were considered with varying values of virial coefficients $k$. \citet{orlov10} obtained a power-law index of $0.7073 \pm 0.0016$ for the lifetime distribution of systems with $k = 0$, which is the same as the initial condition we adopt in our three-body interactions. Therefore, it appears that \citet{shevchenko10}\textquotesingle s analysis of LF1-type L\'evy flights and the resulting heavy tailed distribution of these flights is indeed general in the sense that they are applicable in both the unequal mass limit, like chaotic cometary systems, and also in systems close to the equal mass limit. This suggests that LF1-type flights are a fundamental property of three-body interactions, unlike LF2-type flights. This is likely due to the fact that LF1-type flight arise from the separatrix between bound and unbound orbits, which is always present in the general three-body regardless of the mass ratio.

Is there a candidate for a different kind of L\'evy flights in the general three-body system in place of LF2-type? As discussed earlier in Section~\ref{ls_rel}, we do observe 2 different kinds of interactions in the L\'evy regime in the $\mathcal{S}$ vs. $\mathcal{L}$ space. However, we hypothesize that the second kind originating from low LF indices are not L\'evy processes.  This is because they go against the idea of L\'evy flights as excursions, where the system has a hierarchical structure. Further investigation will be needed to understand these interactions.

\begin{figure}
\includegraphics[width=\columnwidth]{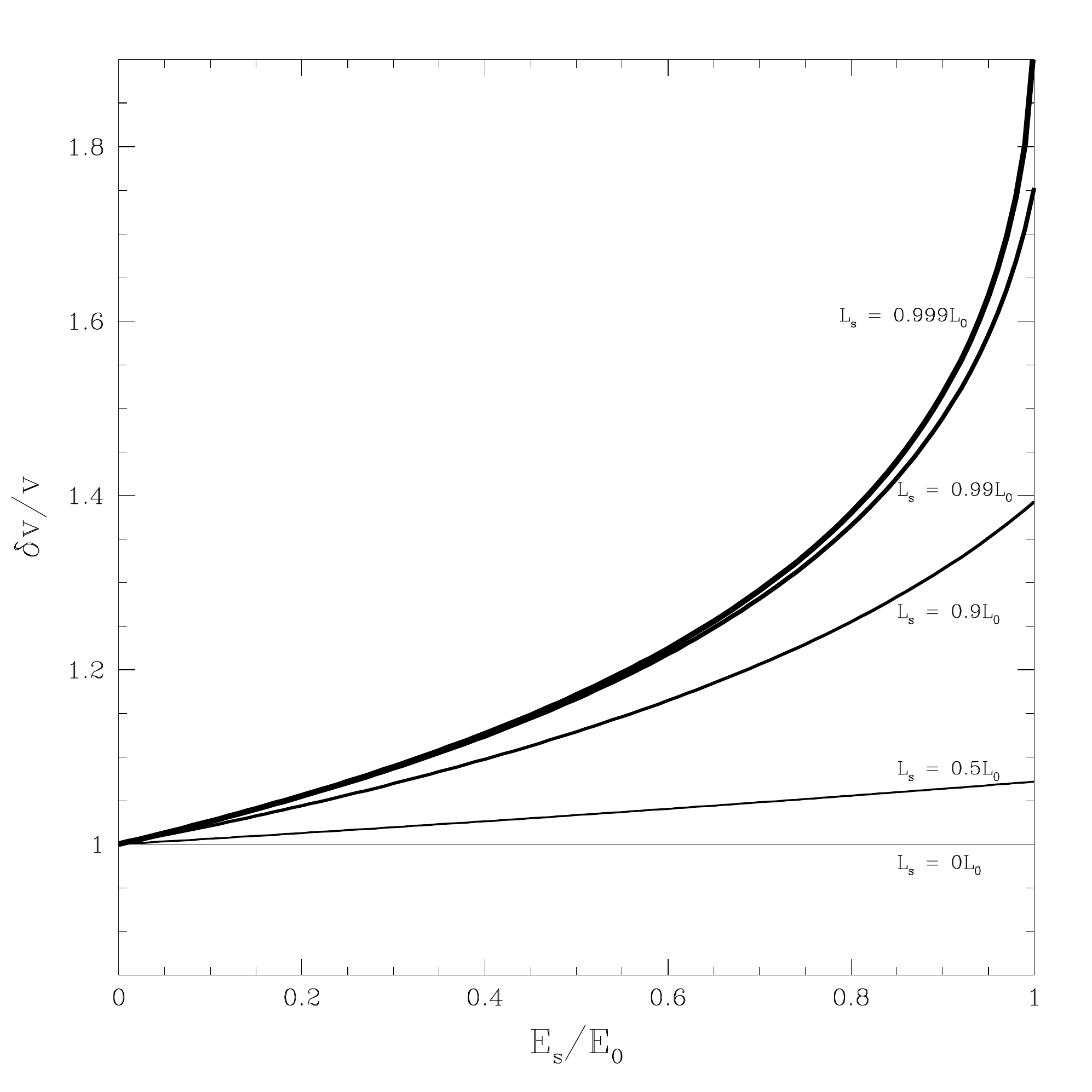}
\caption{Fractional change in velocity $\delta{v}/v$ per pericenter passage as a function of orbital energy and total interaction angular momentum in the temporary single.}
\label{fig:fig5}
\end{figure}

\subsection{What is the physical origin of the long-lived power-law tail in the cumulative lifetime distributions?} \label{tail}

\begin{figure*}
        \centering
        \begin{subfigure}[b]{0.475\textwidth}
            \centering
            \includegraphics[width=\textwidth]{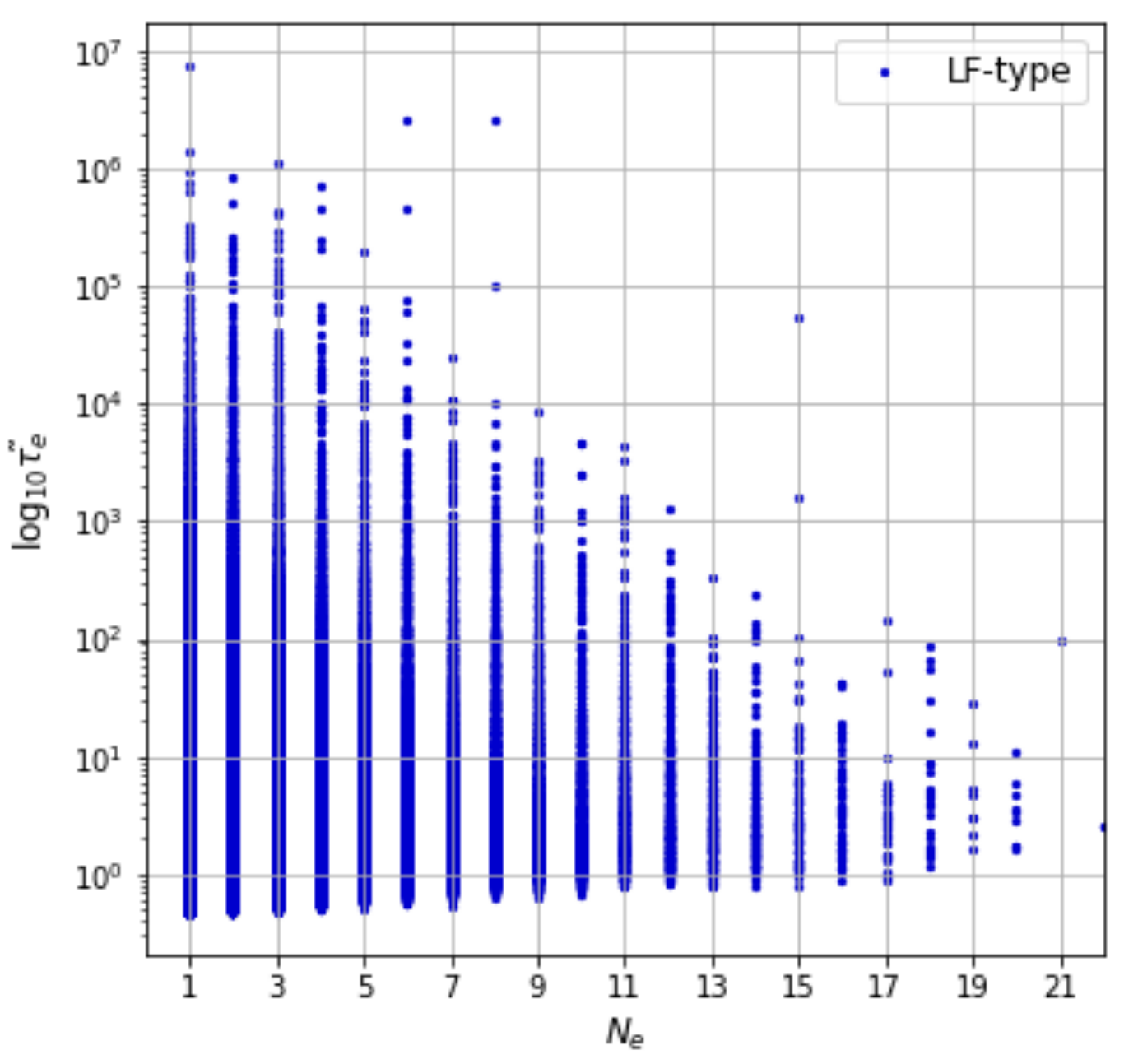}
            \caption{Plot of log of average excursion length ($\log_{10}\tilde{\tau_{e}}$) versus number of excursions ($N_{e}$) for L\'evy flight interactions (LF-type) for 15,15 and 15$M\textsubscript{\(\odot\)}$ system.}   
            \label{fig:phys1}
        \end{subfigure}
        \hfill
        \begin{subfigure}[b]{0.475\textwidth}  
            \centering 
            \includegraphics[width=\textwidth]{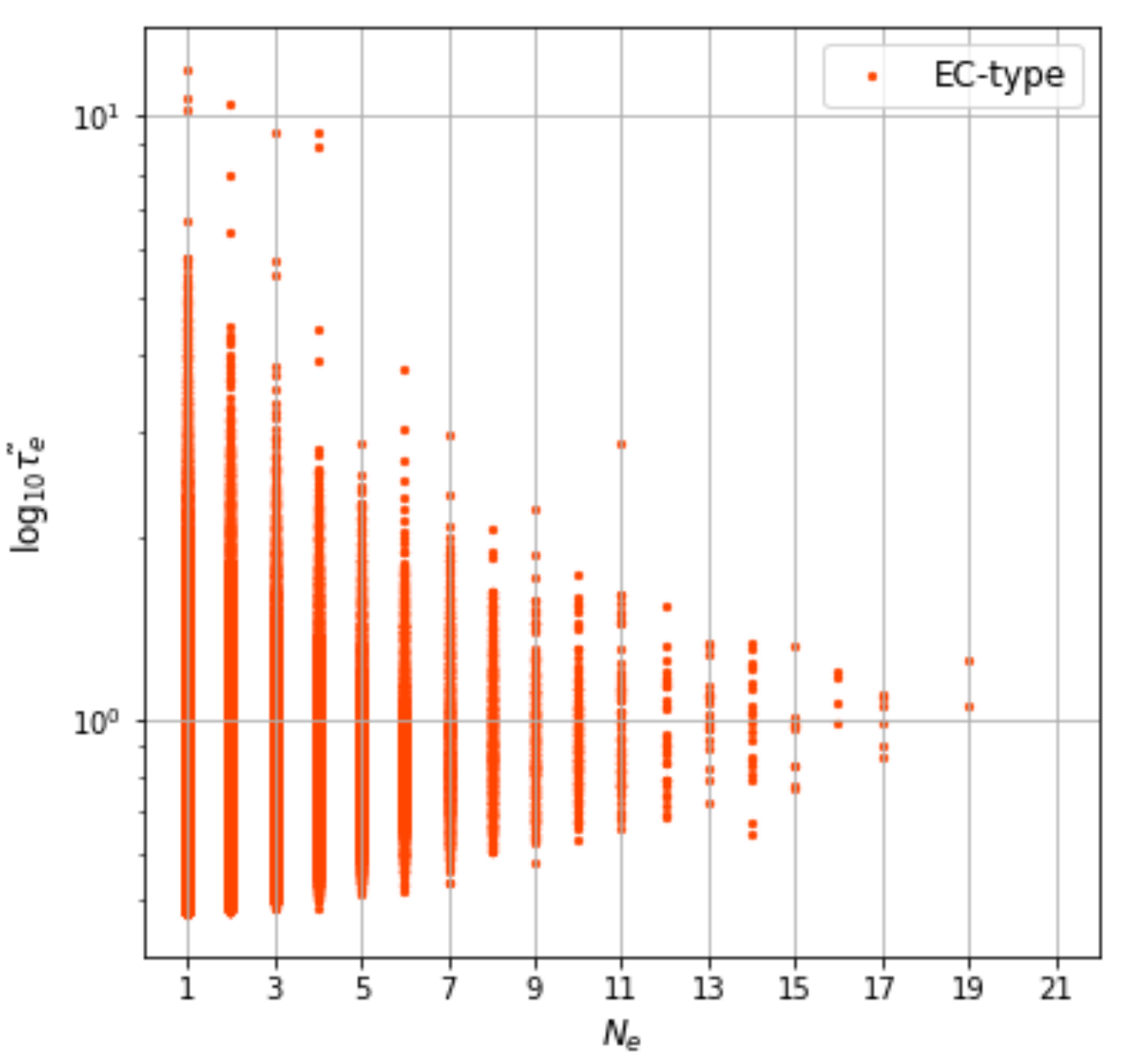}
            \caption{Plot of log of average excursion length ($\log_{10}\tilde{\tau_{e}}$) versus number of excursions ($N_{e}$) for EC-type interactions for 15,15 and 15$M\textsubscript{\(\odot\)}$ system.}    
            \label{fig:phys2}
        \end{subfigure}
        \vskip\baselineskip
        \begin{subfigure}[b]{0.475\textwidth}   
            \centering 
            \includegraphics[width=\textwidth]{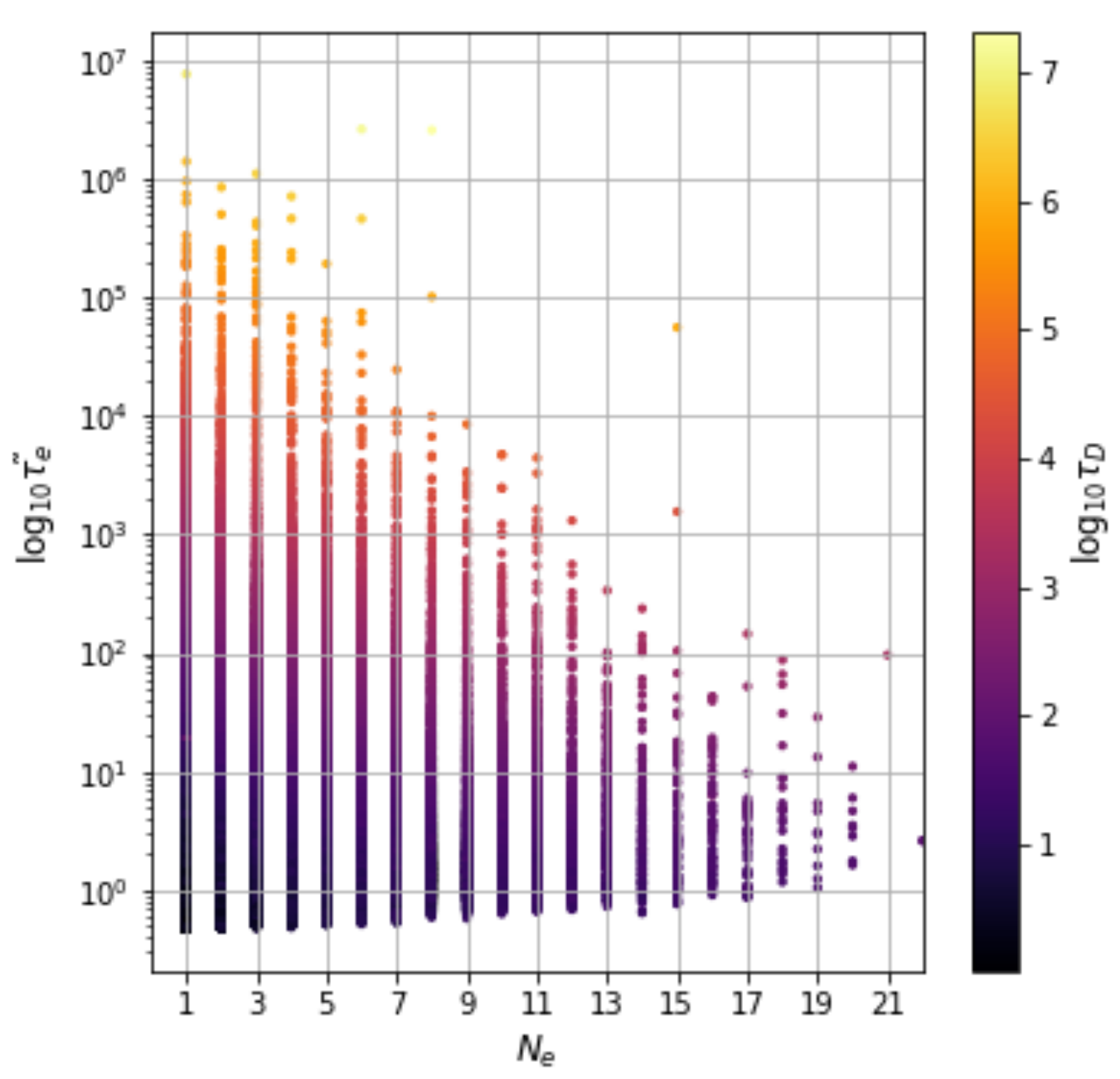}
            \caption{Plot of log of average excursion length ($\log_{10}\tilde{\tau_{e}}$) versus number of excursions ($N_{e}$) for all interactions color coded according to lifetime $\tau_{d}$ for 15,15 and 15$M\textsubscript{\(\odot\)}$ system.}    
            \label{fig:phys3}
        \end{subfigure}
        \hfill
        \begin{subfigure}[b]{0.475\textwidth}   
            \centering 
            \includegraphics[width=\textwidth]{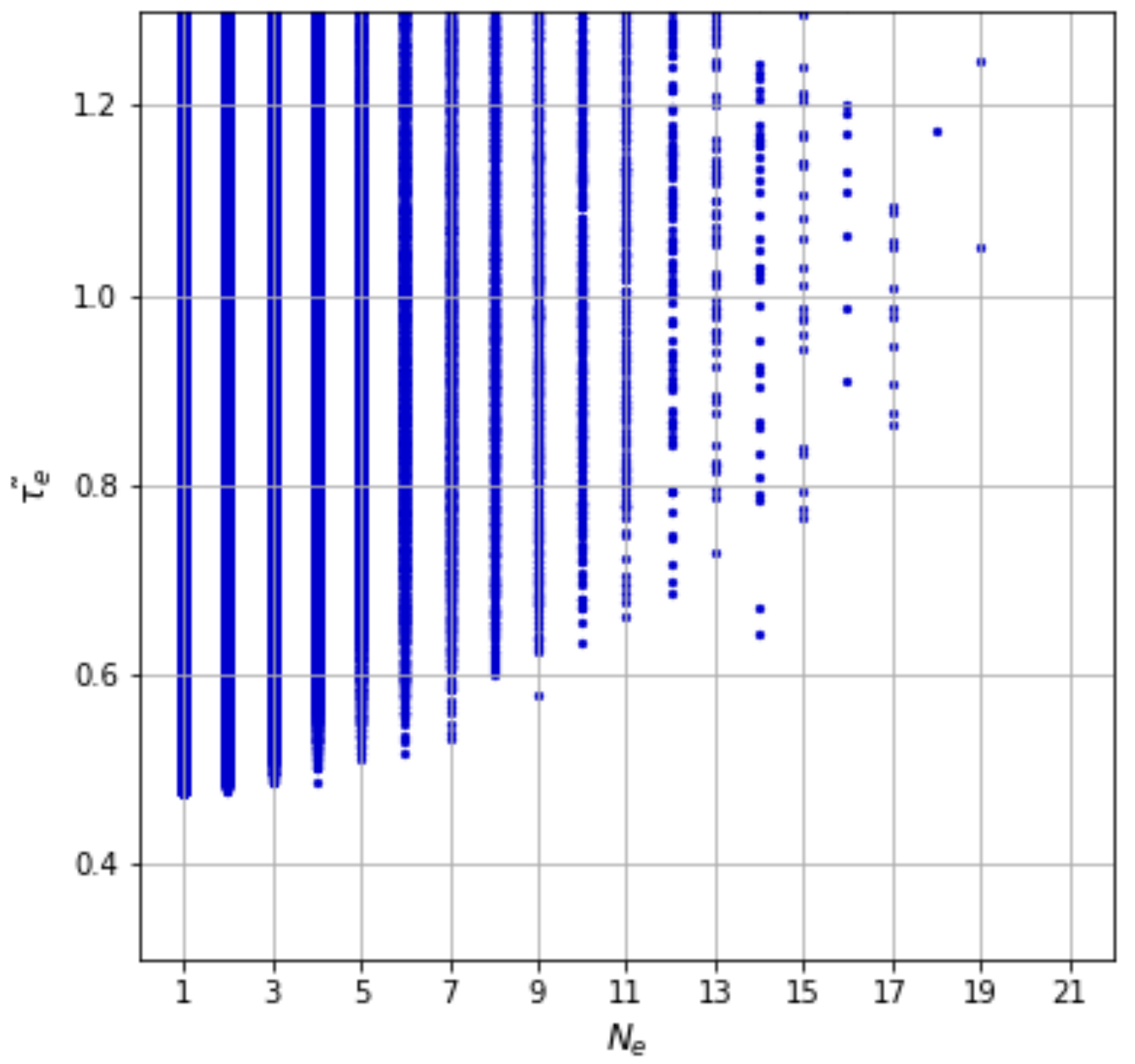}
            \caption{Zoomed in plot (lower bound) of average excursion length ($\tilde{\tau_{e}}$) versus number of excursions ($N_{e}$) for all interactions color coded according to lifetime $\tau_{d}$ for 15,15 and 15$M\textsubscript{\(\odot\)}$ system.}
            \label{fig:phys4}
        \end{subfigure}
        \label{fig:phys_tot}
\caption{Relation between duration of excursion $\tau_{e}$ and number of excursions $N_{e}$ for 15,15,15$M\textsubscript{\(\odot\)}$ system.}
    \end{figure*}

Based on our earlier discussion regarding L\'evy flights and our results regarding the tails of the cumulative lifetime distributions, we know that L\'evy flights are responsible for the long-lived algebraic tail. Based on the homology radius maps for L\'evy flight interactions, we now have a more concrete understanding of the nature of these L\'evy flights as long excursions or as a series of long excursions.  But is it the compounded number of long excursions or a single very long excursion that is governing the long-lived tail at very long interaction times?  What underlying physical mechanism is at play in producing the power-law tails?

To gain physical insight into the answer to this question, consider drawing a parallel between the concept of two-body relaxation in large-N self-gravitating systems and prolonged excursions in the three-body problem.  The former is roughly defined as the time for deflections $\delta v$ in a particle's velocity $v$ to compound sufficiently to change its original velocity by of order itself, or $\delta v$/$v \sim 1$. \citep[e.g.][]{binney87}.  

Now, consider a temporary hierarchy in a chaotically interacting three-body system in which one of the particles is on a prolonged, wide orbit with orbital separation $a_{\rm s}$ and eccentricity $e_{\rm s}$ (relative to the center of mass of the binary, and $a_{\rm s} \gg a_{\rm B}$).  To compute the change in velocity $\delta v$ at closest approach or pericenter, we compute the impulse by taking the product of the (instantaneous) acceleration induced on the particle and the typical time spent at this point (approximating the binary as a point particle in this simple calculation):

\begin{ceqn}
\begin{align}
\label{eqn:dv}
{\delta}v = \Big( \frac{Gm_{\rm B}}{b^2} \Big) \Big( \frac{2{\alpha}b}{v_{\rm p}} \Big),
\end{align}
\end{ceqn}

where $\alpha$ is a free scaling parameter of order unity, m$_{\rm B}$ is the binary mass, b is the distance at closest approach (i.e., pericenter) or $b =r_{\rm p} = a_{\rm s}(1-e_{\rm s})$ and $v_{\rm p}$ is the velocity at pericenter:
\begin{ceqn}
\begin{align}
\label{eqn:vp}
v_{\rm p} = \Big( \frac{Gm_{\rm B}(1+e_{\rm s})}{a_{\rm s}(1-e_{\rm s})} \Big)^{1/2} = \Big( \frac{2E_{\rm B}(1+e_{\rm s})}{m_{\rm s}(1-e_{\rm s})} \Big)^{1/2}.
\end{align}
\end{ceqn}

Plugging the above into Equation~\ref{eqn:dv} and re-arranging, we obtain:

\begin{ceqn}
\begin{align}
\label{eqn:dv2}
\frac{{\delta}v}{v_{\rm p}} = \frac{2\alpha}{1 + (1 - L_{\rm s}^2E_{\rm s}/(L_{\rm 0}^2E_{\rm 0}))},
\end{align}
\end{ceqn}

where $E_{\rm s}$ = $|E_{\rm s}|$ and $L_{\rm s} = |\vec{L_{\rm s}}|$, and also $E_{\rm 0}$ and $L_{\rm 0}$ denote, respectively, the total energy and angular momentum of the interaction.  Since we adopt zero impact parameter, a relative velocity at infinity of zero and an initially circular orbit for the binary, these reduce simply to $E_{\rm 0} = Gm_{\rm s}m_{\rm B}\text{/}(2a_{\rm 0}$) and $L_{\rm 0} =\mu\sqrt{GMa_{\rm 0}}$, where $\mu = m_{\rm s}m_{\rm B}\text{/}(m_{\rm s} + m_{\rm B}$) is the reduced mass, $M= m_{\rm s} + m_{\rm B}$ is the total system mass and $a_{\rm 0}$ is the initial binary orbital separation.

Figure~\ref{fig:fig5} shows the fractional change in the single star velocity $\delta v$ (divided by the initial pericenter velocity) per pericenter passage as a function of the orbital energy of the temporary single (normalized by the total encounter energy). The different line thicknesses correspond to different assumptions for the fraction of the total interaction angular momentum in the temporary single star orbit.  We set $\alpha =$ 1 for illustrative purposes, but caution that the scale on the y-axis should be scaled by some unknown factor (due to the poor approximation in Equation~\ref{eqn:dv}) brought about by assuming that the velocity deflection is induced instantaneously at pericenter). In Figure~\ref{fig:fig5}, we clearly see that more loosely bound temporary singles experience weaker interactions at pericenter, requiring additional such interactions to induce a significant change to the outer orbit of the triple.  Only in the low angular momentum regime can the exchange be strong (since the pericenter distance is smaller).  Qualitatively, this suggests that long or prolonged excursions of a loosely bound temporary single particle are most likely to be followed by additional (albeit slightly reduced) prolonged excursions, relative to initially shorter excursions.  This could help to explain the power-law tail observed in the cumulative lifetime distributions of disintegrating three-body systems \citep[e.g.][]{leigh16,leigh18} and the very nature of L\'evy flights.    

This is precisely what we see in Figure~\ref{fig:phys1}. These figures plot the total number of excursions ($N_{e}$) in the evolution of a three-body system against the average duration of an excursion ($\tilde{\tau_{e}}$) in units of $\tau\textsubscript{cr}$. The average duration of an excursion in a system is defined as the total amount of time spent in the hierarchical state divided by the number of excursions $N_{e}$. Looking at these figures, we mainly see 2 patterns at the lower and upper bounds of the distribution of data points. At the lower-bound of these plots, which is shown in more detail in Figure~\ref{fig:phys4}, as the average length of the excursion increases, the number of excursions also increases, which is in line with our predictions and reasoning discussed earlier in this section. Interestingly, this lower bound behavior is found not only in L\'evy flights but also in EC-type interactions as seen in Figure~\ref{fig:phys1} and Figure~\ref{fig:phys2} respectively. This is indicative of a deeper significance, that is, a result fundamental to the nature of the time evolution of three-body systems, namely a \textit{three-body relaxation mechanism}.

However, there is a small caveat: at the upper bounds of these figures, the opposite behavior occurring. As the number of excursions increases, the average duration of the excursions is going down. The reason for this 'opposite' behavior is due to the finite lifetimes of our three-body systems, since they eventually eject a particle resulting in a termination of the interaction. As seen in Figure~\ref{fig:phys3}, the upper edges of this relation are occupied by longer lifetime $\tau_{D}$ interactions.  Based on the lifetime distributions seen earlier, we know that the fraction of systems with a total interaction lifetime of length $\tau_{D}$ decreases as $\tau_{D}$ increases. Therefore, we are lacking simulations of three-body interactions in the top-right regions of these relations, since the top-right region corresponds to interactions with longer lifetimes. To summarize, we have 2 competing effects in these figures: (i) the three-body relaxation mechanism where longer initial excursions are naturally accompanied by a larger number of excursions, which dominates the lower bounds of these relations, and (ii) a low probability of very long-lived systems, which contributes to under-populating the upper bounds of these relations for large numbers of excursions. 

Therefore, to understand the relation between the length of excursions and the number of excursions identified in Figure~\ref{fig:phys1}, we need only to focus on the lower bounds of these relations. This confirms our earlier hypothesis, namely that as the duration of the initial excursion increases, the total number of excursions also increases.

\subsection{Understanding General Three-body Lifetime Distributions} 
\label{decompose}

In Section~\ref{results}, we discussed the different kinds of three-body interactions, namely EC-type and L\'evy flight interactions. We also discuss in great detail the power-law and exponential models governing the lifetime distributions of these interactions. We find that EC-type interactions, that is interactions in the ergodic limit, follow an exponential decay in the lifetime distribution similar to what has been found in radioactive decay \citep{leigh16,ibragimov18}. On the other hand, the L\'evy flight interactions, as identified in this paper, follow power-law/algebraic lifetime distributions that correspond to the tail of the three-body system lifetime distribution. However, we were able to study these different interactions and distributions by developing metrics, the Scramble Density $\mathcal{S}$ and the LF index $\mathcal{L}$, and then separating the different interactions through methods we developed in Section~\ref{results}. In order to quantify and characterize the lifetime distributions more directly, we introduce a new physically-motivated model for the three-body problem. We first describe two previous models for the lifetime of the general three-body problem. 

\citet{ibragimov18} used the model
\begin{ceqn}
\begin{align}
f(\log(\tau_{D})) = \frac{1}{C} \exp{-\frac{1}{2}(\frac{\log(\tau_{D}/t_{k,0})}{\sigma/\tau_0})^2} 
\end{align}
\end{ceqn}
where $C = \sqrt{2\pi} \cdot \frac{\sigma}{\tau_0}$. The parameter $\tau_{0}$ denotes the right-ward shift of the Gaussian curve, $\sigma$ denotes the widths of the distributions and $\tau_{k,1/2}$ denotes the half-lives. As discussed in \citet{ibragimov18}, the motivation behind this model is that the lifetime distributions resemble Gaussian curves on logarithmic time scales, or $\log_{10}\tau_{D}$. However, understanding the distribution can become more complicated when dealing with and trying to fit super-posed Gaussian curves. 

On the other hand, \citet{leigh16} used the model
\begin{ceqn}
\begin{align}
f(\tau) = \alpha e^{(\tau - \tau_{0,a})/\tau_{1/2,a}} + \coth{((\tau - \tau_{0,b})/\tau_{1/2,b})} + \beta 
\end{align}
\end{ceqn}
where $\alpha, \tau_{0,a}, \tau_{1/2,a}, \tau_{0,b}, \tau_{1/2,b}$ and $\beta$ are free parameters. This model is better than the previous model in the sense that it specifically identifies the initial exponential decay in the lifetime distribution corresponding to the EC-type interactions. Furthermore, as discussed in \citet{leigh16}, the $\beta$ term denotes the L\'evy flights with the $\coth$ term serving as a smooth transition between these two extreme regimes in the full cumulative lifetime distributions. However, based on our discussion of L\'evy flights, we know that L\'evy flights that populate the tail follow a power-law distribution on their own. Hence, we adopt a slightly modified model, as described below.

\begin{table*}
\caption{Median parameter values with 1$\sigma$ uncertainties for \citet{leigh16} model fitted to three-body lifetime distribution.}
\label{table:decomp2}
\centering
\begin{tabular}{ccccccccc}
\hline
Masses ($M\textsubscript{\(\odot\)}$)  & $\alpha$ & $\tau_{0,a}$ & $\tau_{1/2,a}$ & $\tau_{0,b}$ & $\tau_{1/2,b}$ & $\beta$ & $\chi^2_{\nu}$ & AIC \\
\hline
15,15,15 & $-0.938 \pm 0.151$ & $-5546 \pm 5561 $ & $37337 \pm 4565 $ & $ -4.327 \pm 0.239$ & $ 14.93 \pm 0.401$ & $0.119 \pm 0.060$ & 1.056 & 14.80 \\

12.5,15,17.5 & $-1.004 \pm 0.148$ & $ -2307 \pm 4524$ & $ 34797 \pm 4080$ & $ -4.221 \pm 0.250$ & $15.66 \pm 0.43 $ & $0.106 \pm 0.067$ & 1.104 & 19.78\\

10,15,20 & $-1.047 \pm 0.101$ & $-1531 \pm 3797$ & $46138 \pm 6893$ & $-1.894 \pm 0.229$ & $13.37 \pm 0.41$ & $0.110 \pm 0.066$ & 1.441 & 45.86 \\
\hline
\end{tabular}
\end{table*}

\begin{table*}
\caption{Median parameter values with 1$\sigma$ uncertainties for our proposed model fitted to three-body lifetime distribution.}
\label{table:decomp1}
\centering
\begin{tabular}{cccccccc}
\hline
Masses ($M\textsubscript{\(\odot\)}$)  & A & $\tau_{0}$ & B & $\alpha$ & $\chi^2_{\nu}$ & AIC \\
\hline

15,15,15 & $0.669\pm 0.007$ & $4.391\pm 0.027$ & $1.375\pm 0.010$ & $0.797\pm 0.001$ & 0.992 & -9.19\\

12.5,15,17.5 & $0.792 \pm 0.005$ & $4.312 \pm 0.028$ & $1.446 \pm 0.005$ & $0.799 \pm 0.001$ & 0.987 & 2.93 \\

10,15,20 & $1.644 \pm 0.019$ & $3.275 \pm 0.037$ & $1.276 \pm 0.009$ & $0.799 \pm 0.001$ & 1.007 & 5.85\\

\hline
\end{tabular}
\end{table*}

It is easy to think of the general lifetime distribution as a superposition of the underlying exponential decay from the EC-type interactions and the power-law distribution from the L\'evy flights. Therefore, it is natural to try to describe the general model as
\begin{ceqn}
\begin{align}
\label{model}
f(\tau_{D}) = A e^{-\tau_{D}/\tau_{0}} + B \tau_{D}^{-\alpha}
\end{align}
\end{ceqn}
where A,B,$\tau_{0}$ and $\alpha$ are all free parameters. $A$ and $B$ are general constants that do not necessarily have any physical meaning but help normalize the total distribution $f(\tau_{D})$, since they facilitate integrating over suitable limits to normalize it to unity. $\tau_{0}$ is the decay parameter corresponding to the characteristic half-life $\tau_{1/2}$ of the EC-type interactions. $\alpha$ is the power-law index of the distribution tail that is composed of L\'evy flights.

If our hypothesized distribution function shown in Equation~\ref{model} is able to properly characterize the entire lifetime distribution, we should be able to extract information about the underlying contributions from EC-type and L\'evy flight interactions through the fitted values of $\tau_{0}$ and the power-law index $\alpha$.  We can then check to see if these best-fit parameters are similar to the parameter values we obtained through our analysis in Section~\ref{results}.

We perform MCMC fitting (\citet{emcee}) using the proposed model in Equation~\ref{model} and the model provided in \citet{leigh16} to fit our lifetime distributions. The median parameter values along with their 1$\sigma$ uncertainties are shown in Table~\ref{table:decomp1} and Table~\ref{table:decomp2}, respectively. We observe that the AIC values for the new model, namely Equation~\ref{model}, are consistently smaller for all the three-body systems under consideration here, when compared to the \citet{leigh16} model. Therefore, our new model is a better fit to the general three-body lifetime distribution. Table~\ref{table:compare_mod} shows the comparison of the values we obtain for the half-life $\tau_{1/2}$ and power-law index $\alpha$ in Table~\ref{table:decomp1} to the values we obtain in Table~\ref{table:power_entire} and Table~\ref{table:all_half_life} based on our earlier analysis. 

For instance, we find that the power-law index $\alpha$ for the 15,15,15 and 12.5,15,17.5 systems obtained from our model are within $2\sigma$ of the values obtained from our earlier analysis. The results are similar for the half-life $\tau_{1/2}$ in the 10,15,20 system. However, in the other cases, we observe substantial discrepancies, especially in the half-lives $\tau_{1/2}$.
A higher degree of discrepancy is expected in $\tau_{1/2}$.  This is because, during the fitting, the value of the decay parameter $\tau_{0}$ obtained by fitting is quite sensitive to the chosen number of histogram bins in the lifetime distribution and the range of $\tau_{D}$ over which the fitting is being done. This is because, as seen in Section~\ref{sssec:levy_tail}, the exponential part of the three-body lifetime distribution occupies a very small part of the entire distribution. Therefore, considering very large ranges of $\tau_{D}$ or small numbers of histogram bins in the initial part increases the fitting sensitivity in $\tau_{0}$. 
The answer converges adopting the range $0 < \tau_{D} < 10^4$ and 200 bins with varying size (smaller bins in the initial part and bigger bins in the tail).

\begin{figure}
\includegraphics[width=\columnwidth]{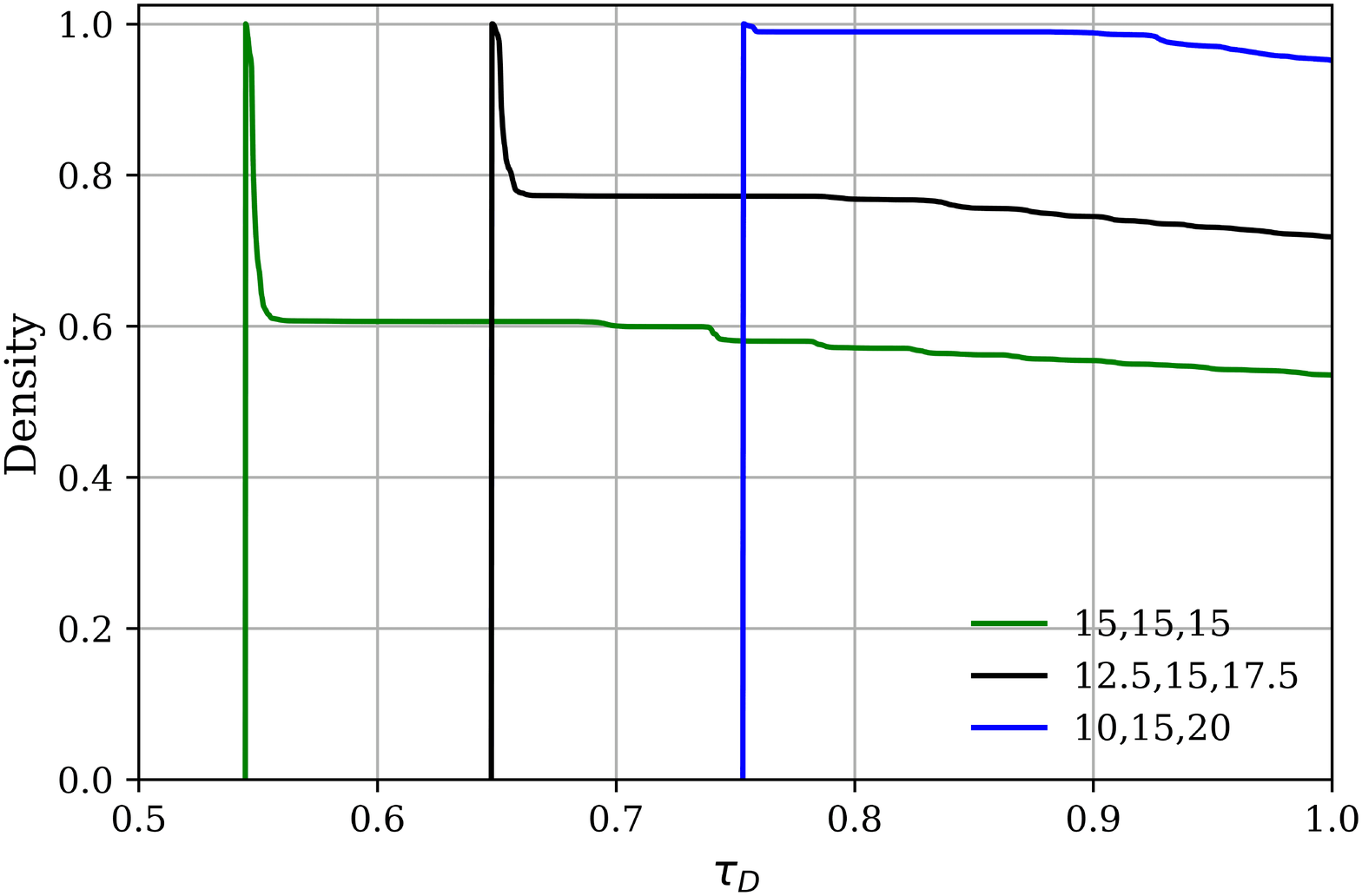}
\caption{Lifetime distributions for all interactions. The small peak at the start of the distributions contains all the ordered interactions that occupy the dark islands as seen in the phase maps in Figure~\ref{fig:fig3}, for example. A smaller range of $\tau_{D}$ is shown on the x-axis to be able to clearly see this peak.}
\label{fig:peak_comp}
\end{figure}

\begin{figure}
\includegraphics[width=\columnwidth]{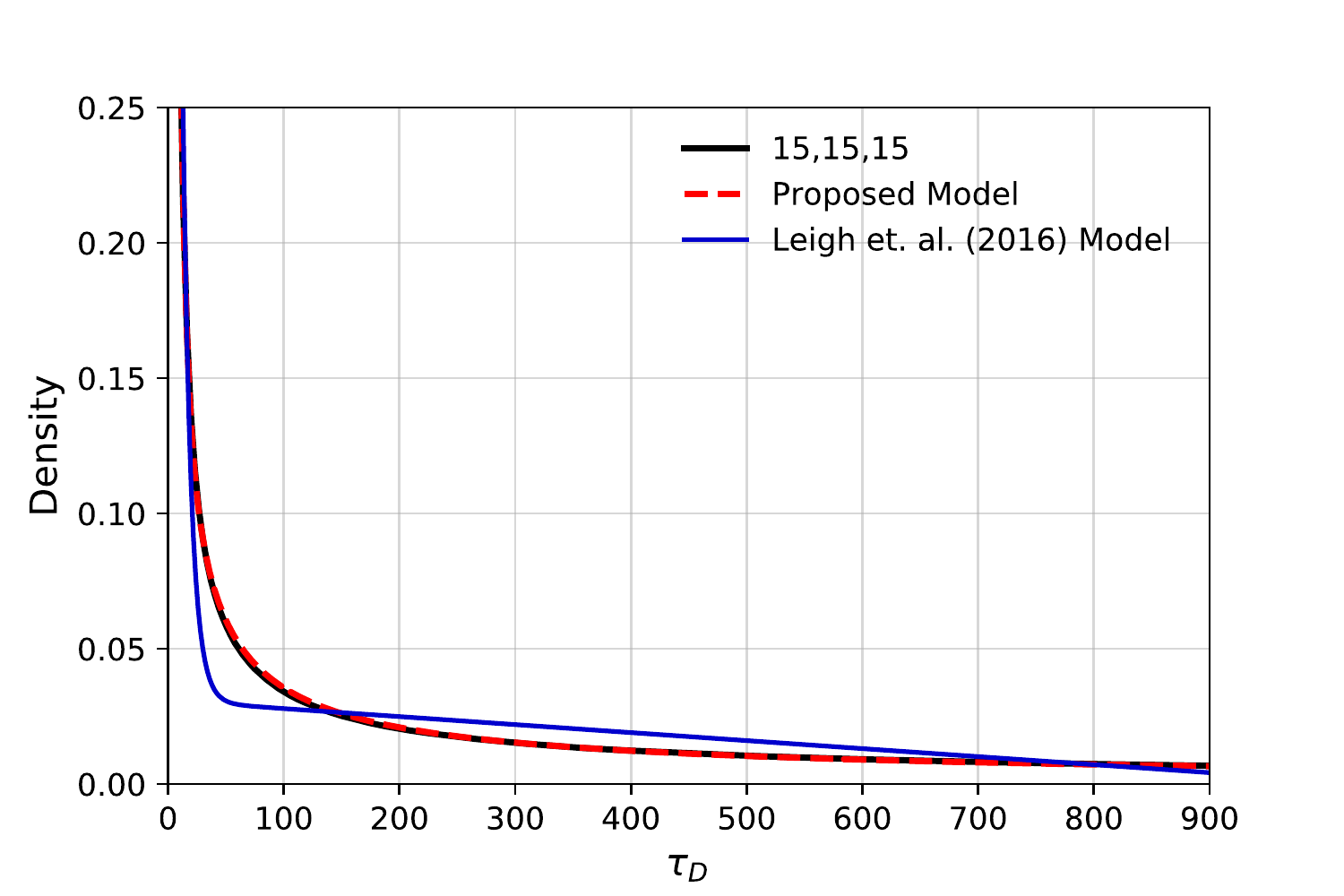}
\caption{Comparison between the new model and the \citet{leigh16} model to the lifetime distribution of the 15,15,15$M\textsubscript{\(\odot\)}$ system. }
\label{fig:final_model}
\end{figure}

\begin{table}
\caption{Comparison of half-life $\tau_{1/2} = \tau_{0}\ln{2}$ and power-law index $\alpha$ values with $1\sigma$ uncertainties calculated from the model proposed in this paper (i.e., New Model) to the earlier obtained values during Power-law tail and Scramble Density Analysis (i.e. Earlier Analysis) }
\label{table:compare_mod}
\centering
\begin{tabular}{ccccc}

\hline
 Masses(M\textsubscript{\(\odot\)}) & Parameter & New Model & Earlier Analysis   \\
\hline
15,15,15 & $\alpha$ & $0.797\pm 0.001$ & $0.783 \pm 0.005$ \\
15,15,15 & $\tau_{1/2}$ & $3.043 \pm 0.027$ & $ 2.580 \pm 0.011 $  \\
12.5,15,17.5 & $\alpha$ & $0.799\pm0.001$ & $0.764 \pm 0.008$ \\
12.5,15,17.5 & $\tau_{1/2}$ & $2.988$ $\pm0.020$ & $ 2.767 \pm 0.014$ \\
10,15,20 & $\alpha$ & $0.799 \pm0.001$ & $0.790 \pm 0.006$\\
 10,15,20 & $\tau_{1/2}$ & $2.270\pm 0.025$ & $ 2.261 \pm 0.011$ \\
\hline
\end{tabular}
\end{table}

Even though this model describes the entire three-body lifetime distribution, there is a small caveat. Throughout our analysis, we have been essentially ignoring the set of ordered interactions; that is, those interactions that terminate quickly due to prompt ejections. These interactions occupy the dark colored islands in the initial condition phase space maps, such as can be seen in Figure~\ref{fig:fig3}. Even though these interactions occupy a major fraction of the phase space, they occupy a very tiny fraction in the time domain. This can be seen in Figure~\ref{fig:peak_comp} where the small, delta-like function peaks in the lifetime distributions. As these peaks occupy a very insignificant fraction of the entire lifetime distribution, they do not significantly influence the exponential decay of the EC-type interactions or the power-law tails. Therefore, in our calculations and analysis, we are largely able to ignore these ordered interactions. They can be very easily isolated via a simple lifetime cut. That is, by only considering long-lived interactions with a total lifetime above some critical value $\tau_{D}$. Describing this delta-function-like peak of the ordered interactions as part of the model is not trivial and will need further investigation in the ordered nature of the three-body problem.

Therefore, we now have a model presented in Equation~\ref{model} that describes general three-body lifetime distributions.  As we have shown, fitting it to our lifetime distributions helps us to extract information about the nature and properties of not only the power-law tails, but also the exponential nature of the EC-type decay interactions.

\subsection{Ratios of Half-Lives and Ejection Probabilities as a Proxy for the Accessible Phase Space Volume}
\label{sssec:half_ratio}

In Section~\ref{sssec:chaos_tails}, we consider three-body interactions in the ergodic limit and find that they follow an exponential lifetime distribution analogous to radioactive decay. We then fit exponential models to the distributions to obtain the half-lives $\tau_{1/2}$ of the different kinds of interactions, as seen in Table~\ref{table:half_life_eject}.  In this section, we compare our computed half-lives and ejection probabilities for different particle masses to different theoretical predictions, namely a naive expectation from energy equipartition-based arguments and the predictions of \citet{stone19} and \citet{kol20}. 

We begin by considering the half-lives for different ejection events in a system.  In the ergodic limit, regardless of the ejection type in a particular three-body system, the half-lives $\tau_{1/2}$ are approximately in the ratio 1:1:1. Note that that this ratio 1:1:1 of half-lives in the ergodic limit is independent of particle masses; that is, this statement is not only true for systems with equal masses, but is also true for systems with unequal masses like 10,15,20$M\textsubscript{\(\odot\)}$. Considering this observation, it is interesting to think of half-lives as more than just a metric for the time scale on which the probability of particle ejection becomes 50$\%$. Could the particle properties, in particular their masses, act as proxies for quantifying or identifying trends in the ratios of accessible phase space volumes for different ejection events? Could there be an underlying physical mechanism in which some of the particle properties decide the probabilities for different ejection types or scenarios to occur? What is the explanation for the equal half-lives for different ejection types in the ergodic limit even in unequal mass systems?  Below, we consider the half-lives for specific ejection events independently, i.e. we consider the time when 50\% of the systems have been disrupted for each of the three possible ejection outcomes.

The inspiration behind this idea comes from Liouville's Theorem \citep{liouville38}. The theorem implies that the total available phase space volume remains constant for self-gravitating systems of point particles, independent of their time evolution. Hence the probability density for accessing any uniform swath of this phase space should also be conserved. In our case, the probability of interest corresponds to observing a specific particle ejection event, or outcome. We naively expect the total phase volume accessible to each particle to be (inversely) proportional to its mass, with less total phase space accessible to more massive particles. 
This stems from the assumption of approximate equipartition of energies in the final distribution of ejection velocities for the escaping star, and the fact that the local escape speed scales linearly with the mean velocity dispersion of the interacting system. This is at least the case for large-N systems of self-gravitating point particles \citep[e.g.][]{binney87}, and it has also been shown to be true for smaller-N systems, at least in a time-averaged sense (see \citealt{valtonen06}, \citealt{leigh16}, and \citealt{leigh18} and \citealt{leigh18b} for more details). Nevertheless, this energy equipartition-based hypothesis has not yet been rigorously tested in the small-N limit.

In the following, we confront the validity of the above energy equipartition-based hypothesis using our simulations, by comparing the half-lives of the cumulative lifetime distributions corresponding to different ejection events (i.e., which of the three masses is ejected).  If the naive hypothesis presented above is correct, our expectation for the ratio of half-lives should be m$_{\rm 1}^{1/2}$:m$_{\rm 2}^{1/2}$:m$_{\rm 3}^{1/2}$. 

\begin{table*}
\caption{Half-lives $\tau_{1/2}$ in units of $\tau\textsubscript{cr}$ for general lifetime distributions. In this context, $\tau_{1/2}$ is when 50\% of the simulations have terminated. Note that this is for all simulation and not just for the ergodic subset. For the equal mass case, Type 0 and 1 correspond to the two 15$M\textsubscript{\(\odot\)}$ particles in the initial binary pair, while Type 2 represents the initial single. }
\label{table:fin_half}
\centering
\begin{tabular}{ccccc}
\hline
 Masses($M\textsubscript{\(\odot\)}$) & Ejection Mass($M\textsubscript{\(\odot\)}$) & $\tau_{1/2}$ & Half-life Ratios & Predicted Ratios (m$_{\rm 1}^{1/2}$:m$_{\rm 2}^{1/2}$:m$_{\rm 3}^{1/2}$) \\
\hline
  & 15 (0) & 3.963 &  &  \\
15,15,15 & 15 (1) & 0.853 & 1 : 0.215 : 0.200 & 1:1:1 \\
  & 15 (2) & 0.793 &  &  \\
 & 12.5 & 4.395 & & \\
12.5,15,17.5 & 15 & 2.374 & 1 : 0.540 : 0.234 & 1 : 1.095 : 1.183 \\
 & 17.5 & 1.031 & & \\
 & 10 & 3.455 & &\\
10,15,20 & 15 & 2.659 & 1 : 0.769 : 0.668 & 1 : 1.224 : 1.414  \\
 & 20 & 2.310 & & \\
\hline
\end{tabular}
\end{table*}

In Table~\ref{table:fin_half}, we show the half-lives for the general three-body lifetime distributions for all interactions (including the prompt, ordered interactions as well) by finding the time $\tau_{D}$ when 50\% of the interactions are terminated; that is, when the distribution takes a value of 0.5. Based on Table~\ref{table:fin_half}, we observe the opposite of what we expect from our hypothesis. That is, we observe that the half-lives of lighter mass bodies are longer than the half-lives of heavier mass bodies. This is because our half-lives are biased by the prompt, ordered interactions. The ejection probability of a lower mass body is higher. Therefore, a higher ejection probability corresponds to a longer half-life. This implies our naive understanding of half-lives is incomplete, and will need further investigation to understand their physical origins.

We also wish to study the ejection probabilities of masses in the purely ergodic limit, not including the subset of ordered interactions. \citet{kol20} discuss the probability distributions of outcomes in the ergodic approximation. The marginalized distributions of energy, average binary energy, angular momentum, eccentricity and escaper probability can also be found. 

For notation purposes, $m_a, m_b$ are the masses in the binary that is leftover after ejection of the escaper mass $m_{s}$. According to \citet{kol20}, the phase space volume (non-normalized probability) for an escaper mass $m_{s}$ in three-body systems with zero total angular momentum, is given by
\begin{ceqn}
\begin{align}
\label{kol_eq}
\sigma_{s} = \frac{1}{3L} \left(\frac{k_{s}}{-2E}\right)^{3/2}
\end{align}
\end{ceqn}
where $E$ is the total energy of the system, $L$ is the total angular momentum of the system, and $k_{s}$ is a dimensionless parameter given by 
\begin{ceqn}
\begin{align}
k_{s} &= \mu_{B}\alpha^2 \\
&= \frac{m_{a} m_{b}}{m_{a} + m_{b} } G^2 m_{a}^2 m_{b}^2 \\
&= \frac{G^2 (m_{a}m_{b})^3}{m_{a} + m_{b}}
\end{align}
\end{ceqn}
For more details about this result and its parameters, we refer the reader to \citet{kol20}. As we are only concerned with the normalized probabilities, we should only concern ourselves with proportionality relations, since the constants cancel. Therefore, 
\begin{ceqn}
\begin{align}
\sigma_{s} &\propto (k_s)^{3/2} \\
&\propto \frac{(m_a m_b)^{9/2}}{(m_a + m_b)^{3/2}}
\end{align}
\end{ceqn}
Running over s = 1,2,3 we can obtain the non-normalized probabilities for the ejection of a mass $m_{s}$ in the ergodic approximation according to \citet{kol20}.

We calculate the ejection probability in the ergodic limit by first isolating all the ergodic interactions using the method we developed in Section~\ref{sssec:scram_results}.  We then calculate the fraction of the total ergodic interactions that result in a particular ejection event, giving us the normalized probability of the ejections based on our simulations. Therefore, the ejection probability $P\textsubscript{s,erg}$ of mass $s$ in the ergodic limit is given by 
\begin{ceqn}
\begin{align}
P\textsubscript{s,erg} = \frac{N\textsubscript{s,erg}}{N\textsubscript{erg}}
\end{align}
\end{ceqn}
where $N\textsubscript{erg}$ is the total number of ergodic interactions among the ensemble of $10^6$ interactions considered and $N\textsubscript{s,erg}$ is the total number of these interactions that eject mass $s$. Assuming Poisson statistics, the uncertainties in our ejection probabilities (\citet{leigh15}) are given by
\begin{ceqn}
\begin{align}
\Delta P\textsubscript{s,erg} = \frac{\sqrt{N\textsubscript{s,erg}}}{N\textsubscript{erg}}
\end{align}
\end{ceqn}

\begin{table*}
\caption{The ejection probability in the ergodic limit compared to the theoretical values predicted by \citet{kol20} and \citet{stone19}. The total number of ergodic realizations ($N\textsubscript{realizations}$) are also shown. The total number of simulations in each set is $10^6$. }
\label{table:eject_prob}
\centering
\begin{tabular}{cccccc}
\hline
 Masses($M\textsubscript{\(\odot\)}$) & Ejection Mass($M\textsubscript{\(\odot\)}$) & $N\textsubscript{realizations}$ & Ejection Probability & \citet{kol20} Probability & \citet{stone19} Probability  \\
\hline
& 15 (0) & 49476 &$0.331\pm 0.002 $  & 0.333 &  0.333 \\
15,15,15 & 15 (1) & 49862 &$0.334\pm0.002$ & 0.333 & 0.333 \\
 & 15 (2) & 50031 &$0.335\pm0.002$ & 0.333 & 0.333 \\
 & 12.5  & 120839 &$0.575\pm 0.002$  & 0.563 & 0.639 \\
12.5,15,17.5 & 15  & 56882 &$0.271\pm0.001$ & 0.279 & 0.247 \\
 & 17.5  & 32434 &$0.154\pm0.001$  & 0.158  & 0.114 \\
 & 10  & 301661 &$0.770\pm 0.001$  & 0.783 & 0.872  \\
10,15,20 & 15  & 65389 &$0.167\pm0.001$ & 0.159 & 0.103   \\
 & 20 & 24876 &$0.063\pm 0.001$ & 0.058 & 0.025 \\
\hline
\end{tabular}
\end{table*}

In Table~\ref{table:eject_prob}, we compare our ejection probabilities in the ergodic limit based on our simulations to the theoretical predictions for the ejection probabilities by \citet{stone19} and \citet{kol20}. We observe that the theoretical predictions for the equal mass system for both formalisms are within $1\sigma$ of the calculated ejection probabilities. In the unequal mass systems, we observe that the calculated ejection probabilities are within $\sim1\%$ of the theoretical predictions for \citet{kol20}. This is highly suggestive of our ergodic assumption being consistent with theoretical formalisms of ergodicity as described in \citet{kol20}.However, \citet{kol20} predictions are not within $3\sigma$ of our calculated probabilities in the unequal mass systems. This suggests that our ergodic subset obtained through a Scramble Density cutoff as described in Section~\ref{sssec:scram_results} will need more refinement, especially in the unequal mass limit. 

Our ejection probabilities are not in good agreement with the the predictions by \citet{stone19}. There are a number of possible reasons for the discrepancy. One of the main reasons is that the ergodic assumption in \citet{stone19}, though similar in nature, is different from our study. The reason why we consider EC-type interactions as ergodic interactions is because these interactions populate the ergodic regions as identified in the initial phase space earlier. \citet{stone19} consider interactions as ergodic if they undergo 2 or more scrambles where a scramble according to \citet{stone19} is periods of time when no pairwise binary exists. However, L\'evy flight interactions could also be considered as ergodic interactions by \citet{stone19} since they undergo scrambles as well. This difference points to the underlying tension in what truly is the ergodic limit in three-body system given the existence of ordered as well as L\'evy flight interactions, especially in the unequal mass limit. Further investigation will be needed to understand the ergodic approximations and corresponding comparisons in the unequal mass limit.

\section{Summary} \label{summary}

In this paper, we introduce a new, robust approach to studying the general gravitational three-body problem. Our goal is to better understand the underlying interactions and physical mechanisms responsible for the observed properties of the three-body lifetime distribution.  To this end, we re-visit the concept of a homology radius in relation to Agekyan-Anosova maps. We introduce two new metrics, namely the Scramble Density $\mathcal{S}$ and LF index $\mathcal{L}$. We demonstrate the utility of these metrics in isolating the different kinds of interactions, namely the ergodic interactions and the L\'evy flight interactions, and their contributions to the lifetime distributions. 

The ergodic interactions that follow an exponential lifetime distribution are called EC-type interactions. Meanwhile the L\'evy flights follow a pure power-law/algebraic distribution. We show conclusive evidence for the fact that the heavy-tailed nature of the three-body lifetime distribution is due to L\'evy flights and the initial exponential drop is due to EC-type interactions. Based on this, we propose a model (Equation~\ref{model}) to fit the lifetime distributions of a general three-body and demonstrate its utility.

Due to the divided phase space nature of the three-body problem, we find both ordered as well as chaotic interactions. The ordered interactions are prompt interactions that lead to the immediate ejection of one of the particles. 
L\'evy flights occupy the ``chaos-order'' border in phase space, giving a halo-like appearance to the islands of ordered interactions in the initial-conditions phase space (see Figure~\ref{fig:fig3}). We further compare the L\'evy flight behaviours in the general three-body problem identified here to the L\'evy flight types, namely LF1-type and LF2-type, established in \citet{shevchenko10} (see Section~\ref{sssec:levy} for a more detailed discussion on this). We show that LF1-type interactions, which are L\'evy flights that appear as sudden jumps in orbital size and period, are the dominant kind of L\'evy flights in a general three-body system.

In relation to the physical origins of L\'evy flights, we introduce a novel mechanism for a \textit{three-body relaxation} process.  We also discuss ejection probabilities of the different masses in the ergodic limit and compare our results to theoretical predictions by \citet{stone19} and \citet{kol20}. We observe that our calculated ejection probabilities are in good agreement with the probabilities predicted by the ergodic formalism of \citet{kol20}. Finally, we illustrate the similarity of the exponential decay distribution of the ergodic EC-type interactions to the radioactive decay observed in nuclear systems. This enables us to further bridge the gap between macro self-gravitating systems and micro nuclear systems: the time evolution can be studied in detail in the former, whereas only the final outcome statistics can be studied in the latter. 

\section*{Acknowledgments}

N.~W.~C.~L. acknowledges the generous support of Fondecyt Iniciac\'ion Grant \#11180005. V. M. acknowledges the generous support of the Jeff Metcalf Fellowship Grant. V. M. is thankful to \textit{Casa Palta} for moral support while the research was conducted. A.A.T. acknowledges support from JSPS KAKENHI Grants \#17H06360 and \#17F17764. We would like to thank  Nicholas C. Stone and Barak Kol for their valuable comments and discussion. We also thank the reviewer for his/her useful comments. Analyses presented in this paper were greatly aided by the following free software packages: \texttt{NumPy} (\citet{numpy}), \texttt{Matplotlib} (\citet{matplotlib}), \texttt{emcee} (\citet{emcee}), \texttt{LMFIT} (\citet{lmfit14}) and \texttt{Jupyter} (\citet{jupyter}). This research has made extensive use of NASA\textquotesingle s Astrophysics Data System and arXiv.

\section*{Data Availability}

The data underlying this article will be shared on request to the authors.

\bibliographystyle{mnras}
\bibliography{main} 
\bsp
\label{lastpage}

\end{document}